\documentclass[review]{cas-dc}

\usepackage{etoolbox} %Used for toggleing between modes.  Specifically here between single column and double column modes.
\providetoggle{double-column}
\settoggle{double-column}{true} %Set True if cas-dc, False if cas-sc

\usepackage{graphicx}% Include figure files
\usepackage{dcolumn}% Align table columns on decimal point
\usepackage{bm}% bold math
\usepackage{wrapfig}
\usepackage{caption}
\usepackage{subcaption}
\usepackage[inline]{enumitem}
\usepackage{placeins} %Gives FloatBarrier
\usepackage[percent]{overpic} %Lets us put text over a figure
\usepackage{xcolor}
\usepackage{url}
\usepackage{lipsum}
\usepackage{cite}
\iftoggle{double-column}{\usepackage[switch,mathlines]{lineno}}{\usepackage[mathlines]{lineno}}
\begin{document}

\shortauthors{D. Southall et~al.} %Shows in footer
\shorttitle{The BEACON Prototype Instrument} %Shows in header

\title[mode=title]{Design and Initial Performance of the Prototype for the BEACON Instrument for Detection of Ultrahigh Energy Particles}

\tnotemark[1]
\tnotetext[1]{This work is supported by NSF Awards \# 2033500, 1752922, 1607555, PHY-2012980 \& DGE-1746045 as well as the Sloan Foundation, the RSCA, the Bill and Linda Frost Fund at the California Polytechnic State University, and NASA (support through JPL and Caltech as well as Award \# 80NSSC18K0231). This work has received financial support from Xunta de Galicia 
(Centro singular de investigaci\'on de Galicia accreditation 2019-2022), 
by European Union ERDF, by the ``Mar\'\i a de Maeztu'' Units of Excellence program MDM-2016-0692, 
the Spanish Research State Agency and from Ministerio de Ciencia e Innovaci\'on PID2019-105544GB-I00 
and RED2018-102661-T (RENATA). We thank the NSF-funded White Mountain Research Station for their support and in particular Steven DeVanzo, Jeremiah Eanes, and all of the staff at Barcroft and Crooked Creek Stations. Computing resources were provided by the University of Chicago Research Computing Center.}

\author[1]{D.~Southall}[orcid=0000-0002-7577-6095]\cormark[1]%orcid 0000-0002-7577-6095
\author[1]{C.~Deaconu}
\author[2]{V.~Decoene}
\author[1]{E.~Oberla}
\author[2]{A.~Zeolla}
\author[3]{J.~Alvarez-Mu\~{n}iz}
\author[2]{A.~Cummings}
\author[1]{Z.~Curtis-Ginsberg}
\author[2]{A.~Hendrick}
\author[1]{K.~Hughes}
\author[2]{R.~Krebs}
\author[1,4]{A.~Ludwig}
\author[5,6]{K.~Mulrey}
\author[7]{S.~Prohira}
\author[5]{W.~Rodrigues~de~Carvalho,~Jr.}
\author[8]{A.~Rodriguez}
\author[4,9]{A.~Romero-Wolf}
\author[10,6]{H.~Schoorlemmer}
\author[1]{A.~G.~Vieregg}
\author[2,8]{S.~A.~Wissel}[orcid=0000-0003-0569-6978]\cormark[2] %orcid 0000-0003-0569-6978
\author[3]{E.~Zas}

\affiliation[1]{organization={Dept. of Physics, Enrico Fermi Institute, Kavli Institute for Cosmological Physics, University of Chicago, Chicago, IL 60637}}
\affiliation[2]{organization={Dept. of Physics, Dept. of Astronomy and Astrophysics, Pennsylvania State University, University Park, PA 16802}}
\affiliation[3]{organization={Instituto Galego de F\'\i sica de Altas Enerx\'\i as IGFAE, Universidade de Santiago de Compostela, 15782 Santiago de Compostela, Spain}}
\affiliation[4]{organization={Jet Propulsion Laboratory, Pasadena, CA 91109, USA}}
\affiliation[5]{organization={Department of Astrophysics/IMAPP, Radboud University, Nijmegen, The Netherlands}}
\affiliation[6]{organization={Nikhef, Science Park Amsterdam, Amsterdam, The Netherlands}}

\affiliation[7]{organization={Dept. of Physics, Center for Cosmology and AstroParticle Physics, The Ohio State University, Columbus, OH 43210}}

\affiliation[8]{organization={Physics Dept., California Polytechnic State University, San Luis Obispo, CA 93407}}

\affiliation[9]{organization={Dept. of Astronomy, California Institute of Technology,  Pasadena, CA 91109, USA}}

\affiliation[10]{organization={Department of High Energy Physic/IMAPP, Radboud University, Nijmegen, The Netherlands}}

\cortext[1]{dsouthall@uchicago.edu}
\cortext[2]{wissel@psu.edu}

\begin{abstract}
The Beamforming Elevated Array for COsmic Neutrinos (BEACON) is a planned neutrino telescope designed to detect radio emission from upgoing air showers generated by ultrahigh energy tau neutrino interactions in the Earth. This detection mechanism provides a measurement of the tau flux of cosmic neutrinos. We have installed an 8-channel prototype instrument at high elevation at Barcroft Field Station, which has been running since 2018, and consists of 4 dual-polarized antennas sensitive between 30-80 MHz, whose signals are filtered, amplified, digitized, and saved to disk using a custom data acquisition system (DAQ). The BEACON prototype is at high elevation to maximize effective volume and uses a directional beamforming trigger to improve rejection of anthropogenic background noise at the trigger level.  Here we discuss the design, construction, and calibration of the BEACON prototype instrument. We also discuss the radio frequency environment observed by the instrument, and categorize the types of events seen by the instrument, including a likely cosmic ray candidate event.
\end{abstract}


\date{\today}

\maketitle

%\tableofcontents

\section{\label{sec:motivation}Introduction}

Cosmic rays have been measured at energies in excess of 100 EeV~\cite{Aab_2020,Abu_Zayyad_2013,ABBASI200953}, above the GZK cutoff~\cite{greisen1966end,zatsepin1966upper}. These measurements may imply the existence of cosmogenic ultrahigh energy (UHE) neutrinos produced by interactions of UHE cosmic rays with cosmic microwave background (CMB) photons~\cite{Berezinsky:1969erk}, or may be consistent with the end of the cosmic ray spectrum~\cite{PierreAuger:2016use,Fang:2013cba}. In either case, measurements or constraints of the neutrino flux at UHE will improve our understanding of cosmic ray accelerators and their cosmological distribution (see. \textit{e.g.} References~\cite{AlvesBatista:2019tlv} and ~\cite{Coleman:2022abf} for recent reviews). In addition to the predicted cosmogenic neutrino flux, recent discoveries of a diffuse flux of astrophysical neutrinos~\cite{IceCube:2013low, IceCube:2020acn,IceCube:2020wum} and a candidate for an extra-galactic source of neutrinos~\cite{doi:10.1126/science.aat1378} create strong motivation for expanding the capabilities of UHE neutrino detectors. Though only electron neutrinos and muon neutrinos are expected to be produced at the sources, flavor oscillations over the astrophysical length scales should result in an observed flavor ratio flux at Earth of 1:1:1~\cite{Choubey:2009jq, Bustamante:2015waa,Pakvasa:2007dc} and any deviations from this could indicate new physics~\cite{Song:2020nfh}. An exclusive measurement of the tau neutrino flux would yield both flux and flavor ratio information for testing both cosmogenic and astrophysical neutrino models~\cite{Bustamante:2019sdb, Xing:2006uk}.  Studying these UHE particles also provides a measurement of interaction cross sections at center-of-mass energies not achievable by current or planned collider experiments, and has the potential to reveal new physics~\cite{Valera:2022ylt, Denton:2020jft, Connolly:2011vc, Esteban:2022uuw}.

\begin{figure*}[htbp]
\centering
\includegraphics[width=\iftoggle{double-column}{1.5\columnwidth}{0.8\textwidth}]{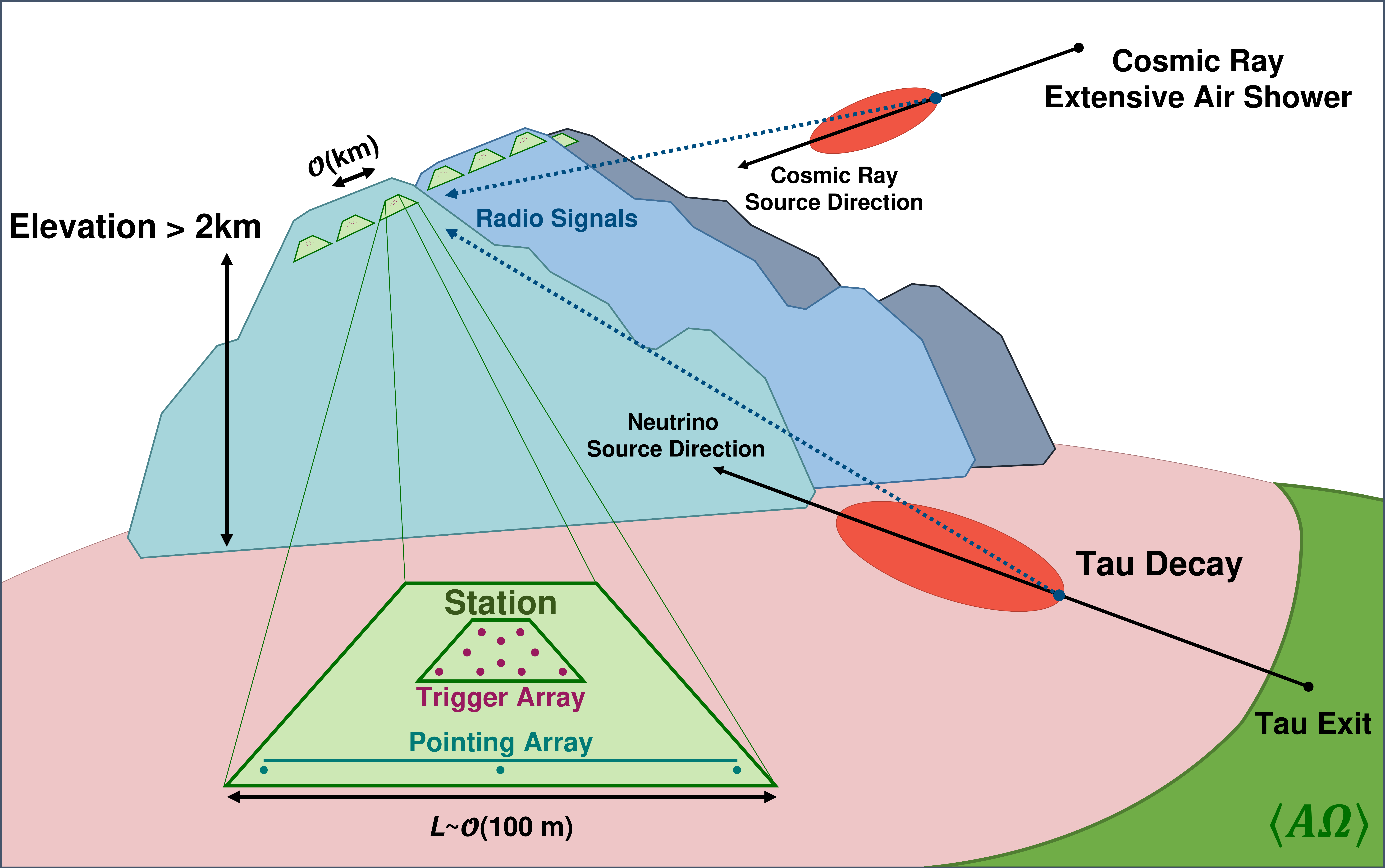}
\caption{A schematic overview of the BEACON concept, adapted from~\cite{wissel2020prospects}. Tau neutrinos interacting in the Earth can produce a tau lepton that escapes into the atmosphere, producing an upgoing air shower upon decay. Radio emission from the air shower may be detected by mountaintop radio stations, each consisting of a small antenna array used for triggering and reconstruction. BEACON stations are also sensitive to emission from cosmic ray-induced air showers, which will come from above the horizon, and may be used for detector characterization.}
\label{fig:concept-schematic}
\end{figure*}

The Beamforming Elevated Array for COsmic Neutrinos (BEACON) concept consists of mountaintop phased radio antennas that are designed for measuring the flux of tau neutrinos above 100 PeV~\cite{wissel2020prospects}.  At these energies, tau neutrinos interacting with the Earth via a charged current interaction can produce a tau lepton boosted enough such that it may escape the Earth and decay in the atmosphere~\cite{Fargion:1999se, Feng:2001ue, Zas:2005zz}.  The tau lepton decay creates an upgoing extensive air shower that will produce an impulsive radio signal. The primary emission mechanism is geomagnetic radiation, a result of the deflection of charges by the Earth's magnetic field, with contributions from Askaryan radiation~\cite{alvarez2012monte}. Air shower radio signatures have been extensively studied by numerous radio experiments (see \textit{e.g.} References~\cite{Schoorlemmer:2015afa, LOPES:2005ipv, Ardouin:2009zp,Schellart:2013bba, monroe2020self, Bezyazeekov:2018yjw, AERA2018} and References~\cite{Schroder:2016hrv,Huege:2016veh} for recent reviews) and have been modelled at accelerator experiments~\cite{Bechtol:2021tyd, T-510:2015pyu}. The probability that a tau lepton will exit the Earth peaks near and below the horizon~\cite{Alvarez_Mu_iz_2018}. This process is shown schematically in Figure~\ref{fig:concept-schematic}.

There are several detector concepts around the world targeting the tau neutrino flux using this Earth-skimming technique, including particle detectors~\cite{Bertou_2002,Romero-Wolf:2020pzh}, imaging Cherenkov and fluorescence telescopes~\cite{Ahnen_2018,Brown:2021tf,Venters_2020}, and radio arrays both on and near mountains~\cite{GRAND:2018iaj, FLIESCHER2012S124, PierreAuger:2015aqe, TAROGE, monroe2020self} and on balloons~\cite{Prechelt_2022,PUEO_white}. See Reference~\cite{Abraham:2022jse} for a recent review. The BEACON concept is distinct for using phased array triggering on a high-elevation mountain. At high elevation, each BEACON station views a large area over which a tau lepton can emerge. The combination of a large prominence and a steerable phased array trigger capable of triggering on events from hundreds of kilometers away provides an optimized detector design for neutrino searches near the horizon. A full-scale BEACON array would consist of $\mathcal{O}$(1000) independent stations, creating a global network of low-cost high-elevation mountaintop radio arrays designed to search for these signals.

Phased array, or interferometric, triggering and reconstruction also offers additional benefits to the BEACON design~\cite{Vieregg_2016}. Directional beams are formed by delaying and summing signals from individual antennas. The trigger is then formed on the coherent sum of the signals from each antenna, which has a higher signal-to-noise ratio (SNR) than the signal from each antenna for true plane-wave air shower signals, thereby lowering the energy threshold of the detector to 100~PeV~\cite{wissel2020prospects} compared to triggering on individual antenna channels.  Additionally, the trigger thresholds on each beam can be dynamically adjusted in response to changes in the local noise environment. These characteristics enhance the trigger's capability to reject anthropogenic radio frequency interference (RFI), which can help maintain sensitivity to the expected diffuse flux while in noisy environments.

A full-scale BEACON array would consist of many stations in various mountain ranges and countries.  Though some sites might have existing infrastructure that can be leveraged, it is not a requirement for a BEACON site.  Stations should therefore be capable of operating autonomously in remote environments at sites with little to no infrastructure.  This means the system must be low-power and operate off-grid using either solar or wind energy.  Stations should also be minimally capable of transmitting monitoring and house-keeping data off-site, with full data transmission desirable to remove the need for retrieval of hard disks.  Finally, such an array must be easy to deploy, robust to weather and wildlife, and cost-effective.  

Development towards the BEACON experiment has been focused on building a prototype. The goals of the prototype study are to evaluate the performance of an interferometric trigger used in this context, and to use the observed cosmic ray flux to measure the \textit{in-situ} expected performance of the full-scale array. As shown in Figure~\ref{fig:concept-schematic}, the prototype is also sensitive to extensive air showers initiated by downgoing cosmic rays.

Though the prototype instrument is not large enough to detect tau neutrinos, we expect to detect cosmic ray air showers with it.  Cosmic ray air showers come from above the horizon, whereas signals from tau neutrinos would come from below the horizon.  The observed rate of cosmic ray events in the prototype instrument presents a \textit{in-situ} validation of the threshold of the instrument. The threshold is an important factor in determining the expected sensitivity to tau neutrinos, allowing us to better predict the sensitivity of the full-scale BEACON experiment in a data-driven way. 

In this paper, we describe the prototype, its goals, and current performance.  Section~\ref{sec:hardware} gives an overview of the BEACON prototype's design, hardware, and implementation.  Section~\ref{sec:analysis} discusses the performance of the array and the phased array trigger. We also present a study of common sources of RFI backgrounds at the prototype site. We also discuss a cosmic-ray-like impulsive event triggered by an RF-only trigger at high-elevation in a noisy environment. In Section~\ref{sec:conclusion}, we place these results in a broader context and discuss future work.

\section{The BEACON Prototype Instrument\label{sec:hardware}}

\begin{figure*}[htbp]
\centering
\includegraphics[width=\textwidth]{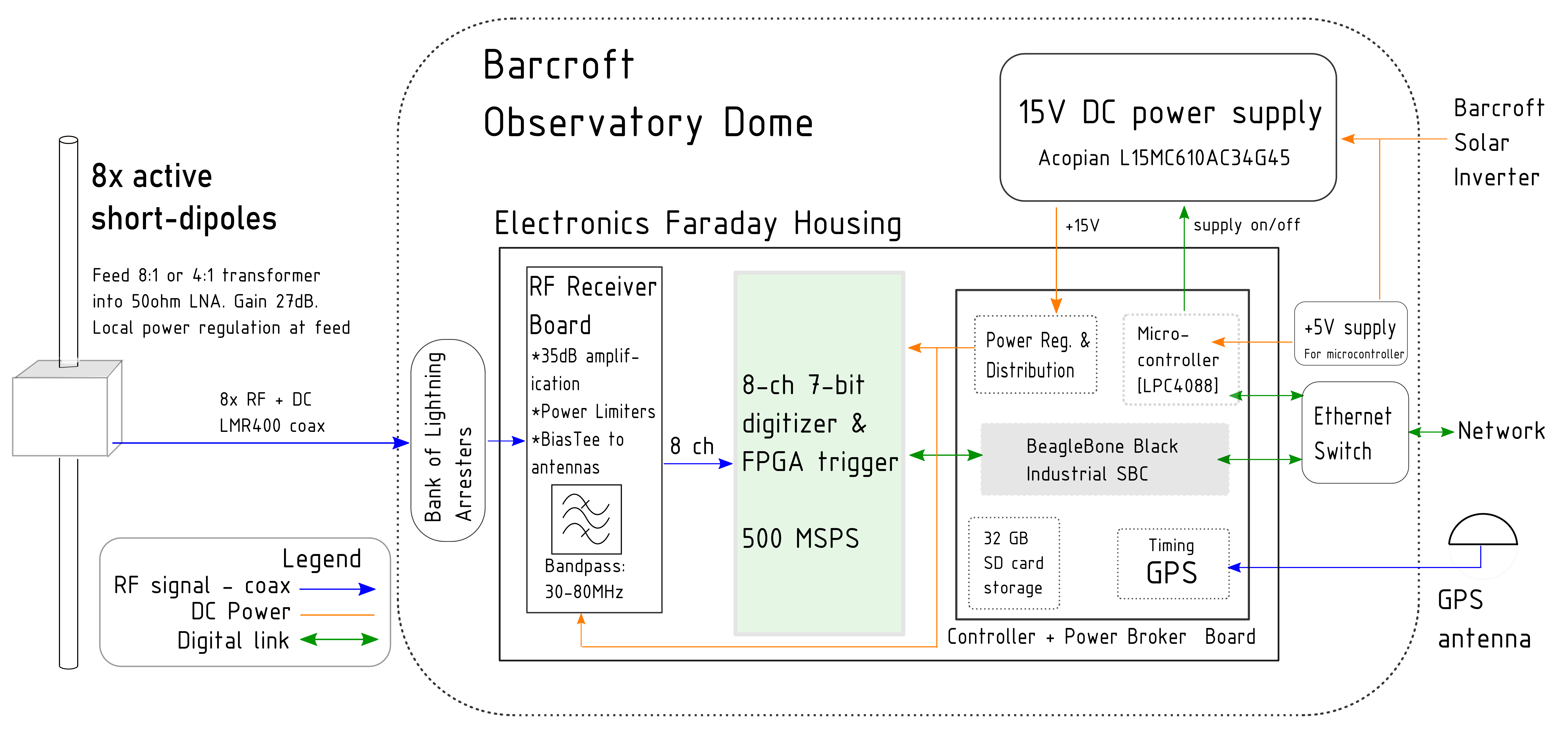}
\caption{Schematic of the BEACON prototype instrument system.}
\label{fig:system-schematic}
\end{figure*}

In 2018, we installed an 8-channel prototype instrument consisting of four dual-polarized antennas and an instrument that amplifies, conditions, and records triggered events.  The system diagram is shown in Figure~\ref{fig:system-schematic}.  The design described here is robust to weather and operating conditions experienced at this remote site, and scalable to larger future deployments.  This section describes the instrument and its site.

\subsection{White Mountain Site\label{sec:wmrc}}

\begin{figure*}[htbp]
\centering
\includegraphics[width=\iftoggle{double-column}{\textwidth}{\textwidth}]{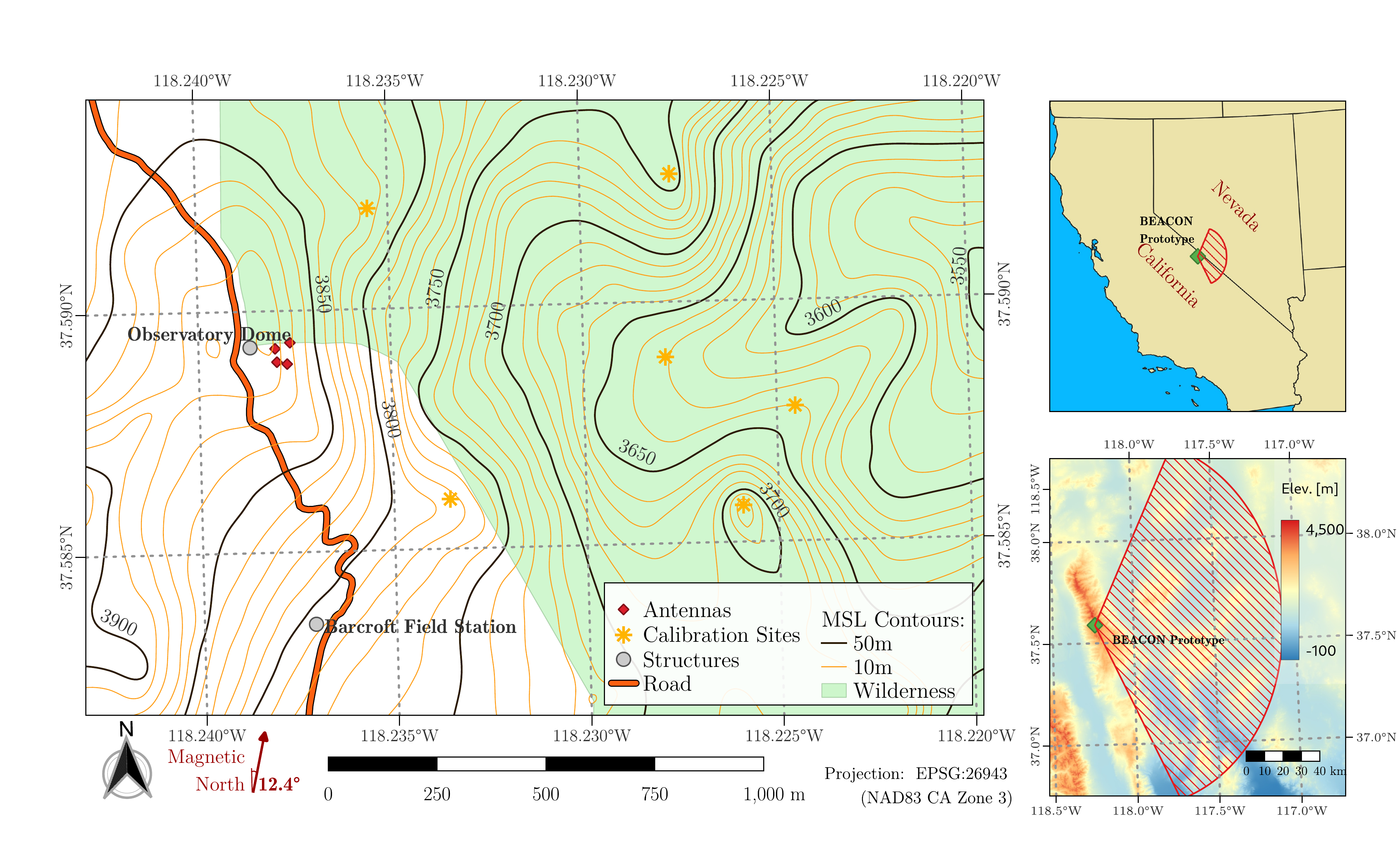}
\caption{Left: A map of the immediate surroundings of the BEACON prototype at the White Mountain Research Station. Electronics are housed in the Observatory Dome.  A scale bar is provided for the local terrain, as well as the direction of magnetic North.  Top Right: A map showing the BEACON prototype's location within California, USA.  Bottom Right: A map showing elevation profile of the region visible to the BEACON prototype.  A cone extended 100 km from the site and spanning $\pm 60^\circ$ of East has been added for reference to illustrate the direction the BEACON prototype faces.}
\label{fig:gis-map}
\end{figure*}

The prototype is located at an altitude of 3.8\,km in the White Mountains of California, near White Mountain Research Center's (WMRC) Barcroft Field Station.  The experiment looks east from the site, overlooking the Fish Lake Valley with the valley floor having an altitude of 1.5~km. Figure~\ref{fig:gis-map} shows the local topography at the site.  The antenna locations are shown in red (and also photographed in Figure~\ref{fig:antennas}), and important structures like Barcroft Field Station and the Observatory Dome are shown in gray.  The Observatory Dome is an enclosed structure with power and network access where our data acquisition system (DAQ) electronics are housed.

The site provides significant infrastructure that is advantageous for BEACON, including road access, a solar-battery hybrid power system, internet access via a microwave relay to Owens Valley Station (which is also operated by WMRC), room and board during deployment, and remote support for the BEACON prototype from WMRC staff.  There are engineering challenges presented by the site that have influenced the design of the prototype instrument: it is only accessible in the summer months, sees wind speeds in excess of 130~km/h, and is located on steep and rocky terrain. Additionally, the permit for the site restricts erecting permanent structures (e.g.\ concrete foundations) under the current agreement with the United States Forest Service.

\iftoggle{double-column}{
%Use if double column
\subsection{Antennas and Mechanical Design}
\begin{figure}[tbph]
\centering
\includegraphics[width=\columnwidth]{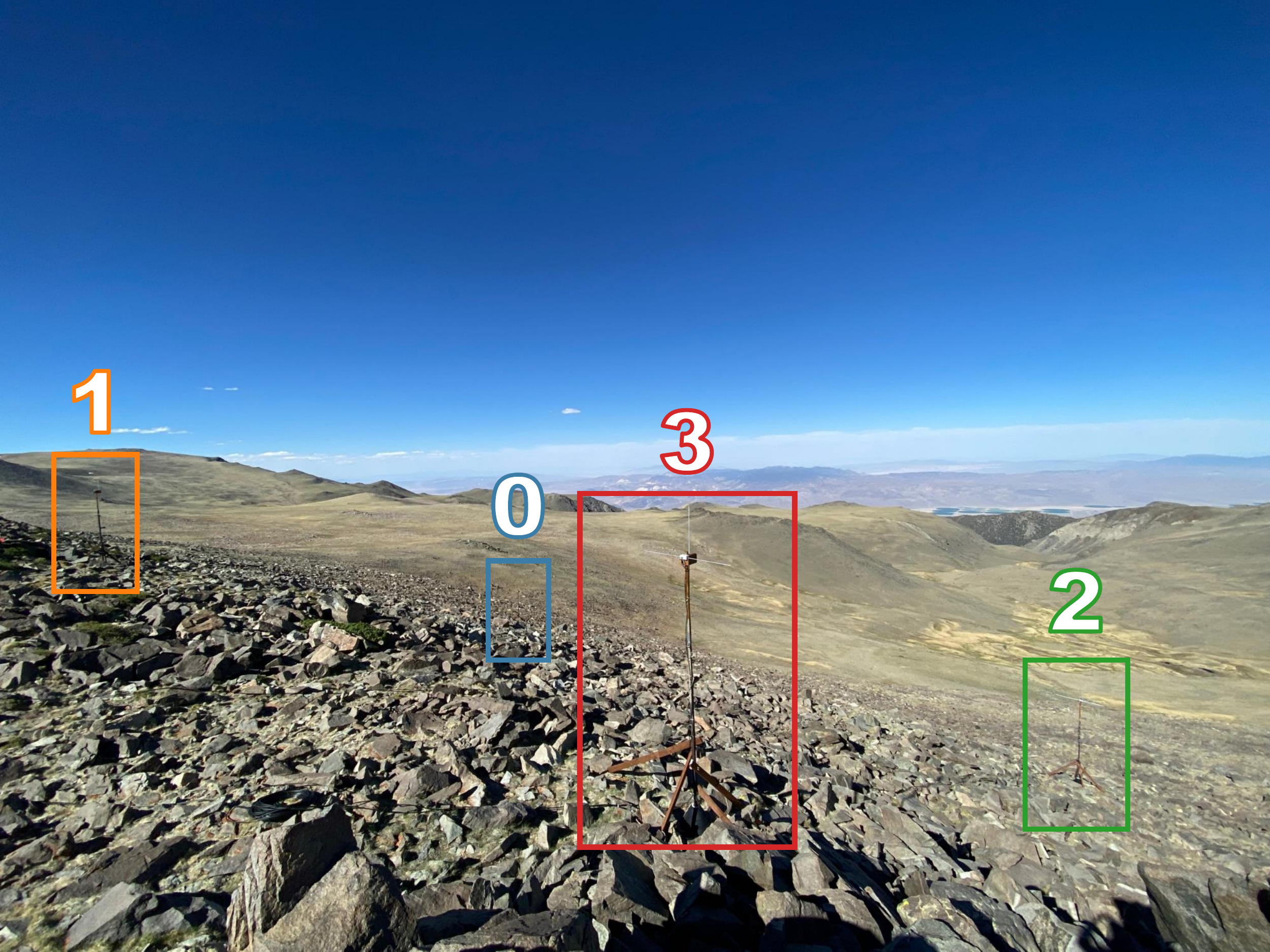}\\
\includegraphics[width=\columnwidth]{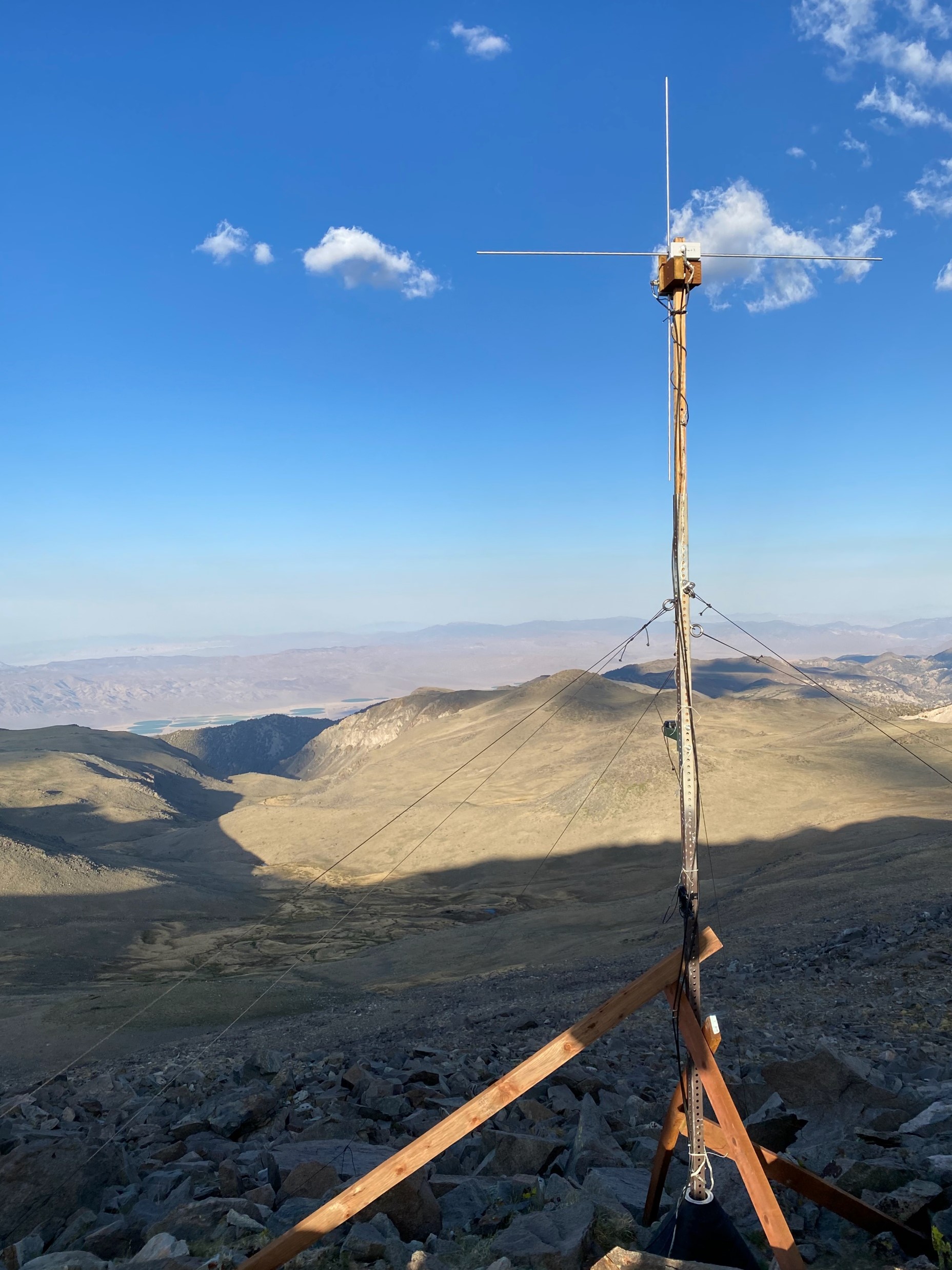}
\caption{Top: The BEACON prototype array consists of four crossed dipoles each with a custom active feed.  The antennas are positioned on a sloped rocky terrain; the HPol (VPol) dipoles are oriented such that their physical extent and gain nulls align in the North-South (Up-Down) direction for maximal sensitivity towards the horizon in the East.  Bottom: Close-up view of Antenna 3 shows the antenna masts with two dipoles and active feeds fed with LMR240 connecting to LMR400 at the base of the antenna. The GPS patch antenna is used for the RTK-based calibration system. The antenna masts are protected against high winds while minimally impacting the local environment using $\sim$33~kg rubber bases, wooden struts, and six guy-lines. All four antennas are elevated $\sim$3.96~m above the ground and pointed toward the horizon to the East. }
\label{fig:antennas}
\end{figure}
}{
%Use if single column
\subsection{Antennas and Mechanical Design}
\begin{figure}[tbph]
\centering
\includegraphics[height=2.5in]{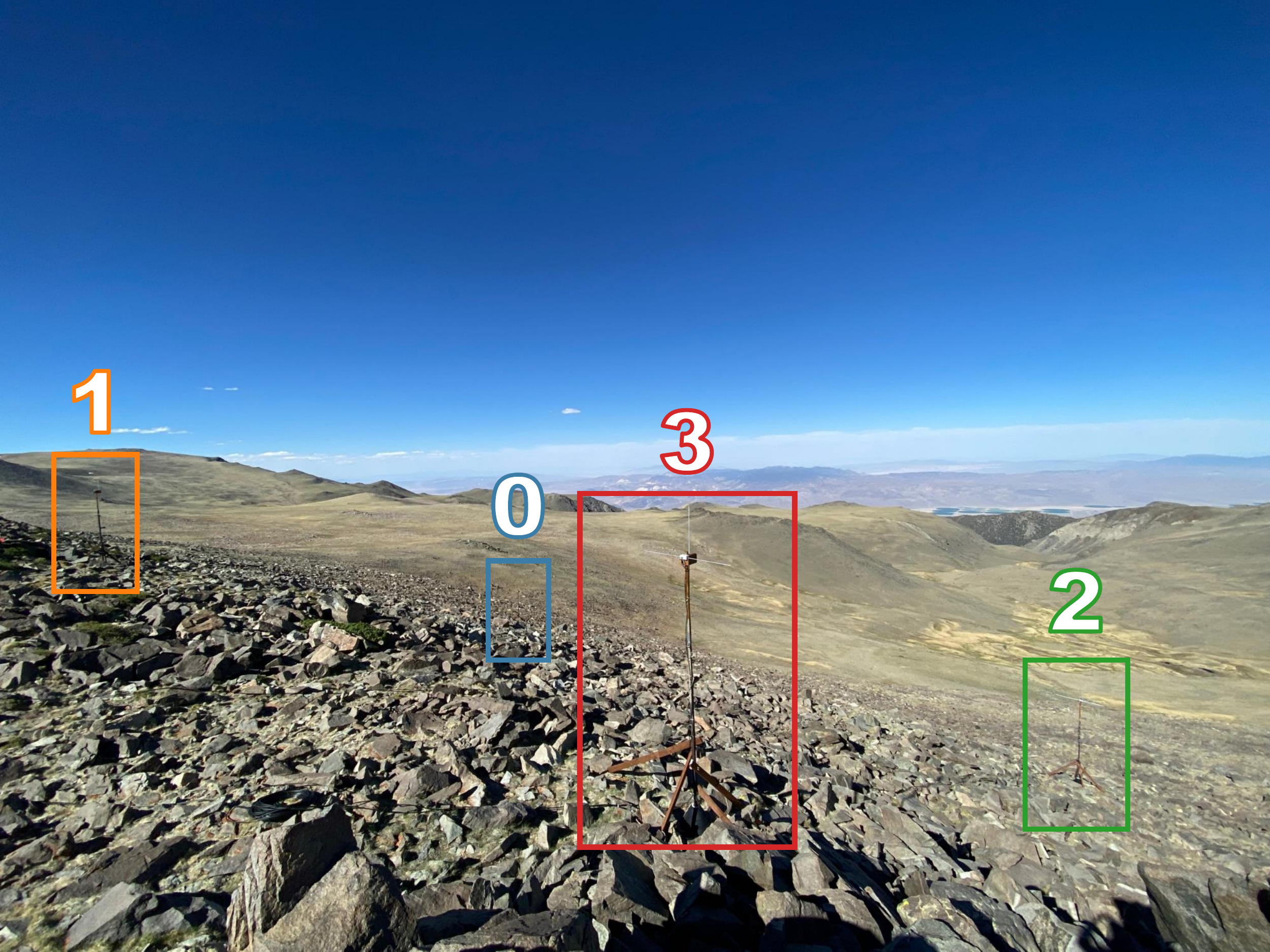}
\includegraphics[height=2.5in]{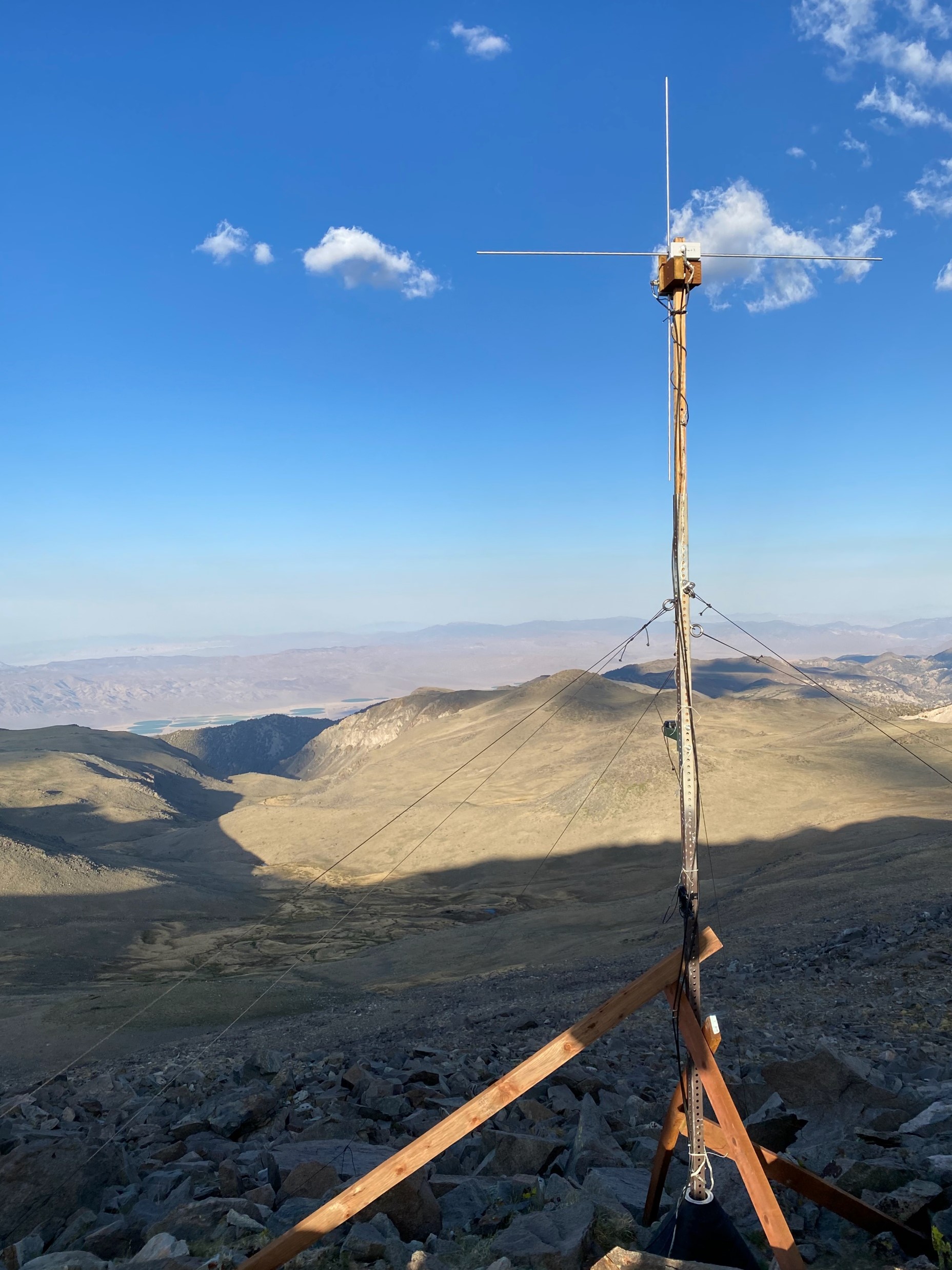}
\caption{Left: The BEACON prototype array consists of four crossed dipoles each with a custom active feed.  The antennas are positioned on a sloped rocky terrain; the HPol (VPol) dipoles are oriented such that their physical extent and gain nulls align in the North-South (Up-Down) direction for maximal sensitivity towards the horizon in the East. Right: Close-up view of Antenna 3 shows the antenna masts with two dipoles and active feeds fed with LMR240 connecting to LMR400 at the base of the antenna. The GPS patch antenna is used for the RTK-based calibration system. The antenna masts are protected against high winds while minimally impacting the local environment using $\sim$33~kg rubber bases, wooden struts, and six guy-lines. All four antennas are elevated $\sim$3.96~m above the ground and pointed toward the horizon to the East. }
\label{fig:antennas}
\end{figure}
}

As the radio emission from air showers is broadband, several bands ranging from 30 to 1200\,MHz can be used to detect them~\cite{wissel2020prospects, V:2019gld}.  Prior to initial deployment of the BEACON prototype, a site survey of RFI was conducted to help make a choice of band~\cite{hughes2019towards}. The antennas chosen for the first implementation of the BEACON prototype were inverted-V cross dipole antennas also used as part of the Long Wavelength Array (LWA) experiment at the Owens Valley Radio Observatory~\cite{LWA_OVRO}.  These antennas were chosen for their sensitivity to 30-80 MHz frequencies, as well as their active balun that includes conversion to a coaxial cable line and amplification of 35 dB~\cite{Ellingson20091421, monroe2020self}. 

\begin{figure*}
    \centering
    \includegraphics[height=0.7\textheight]{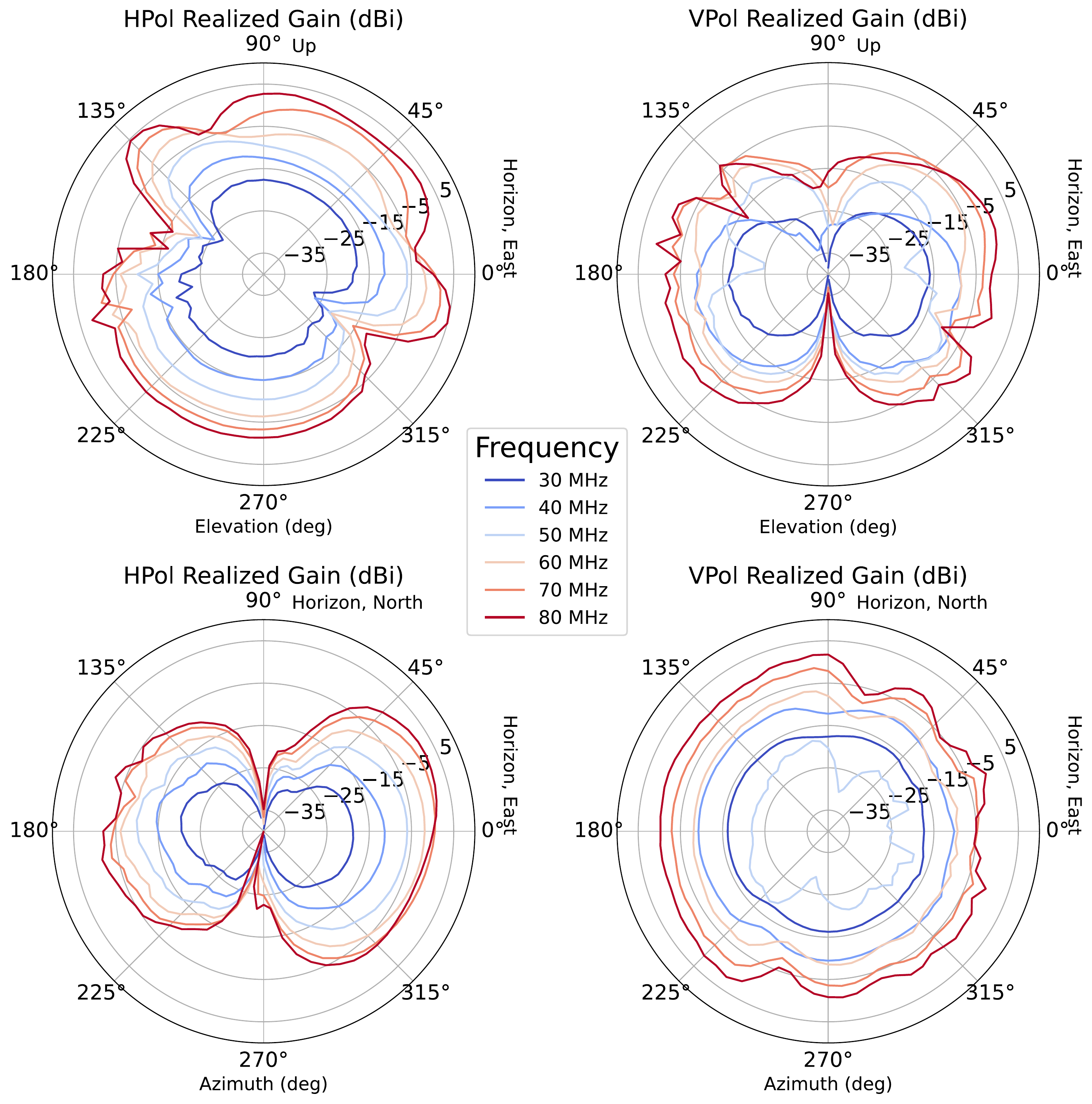}
    \caption{The realized gain of the crossed dipoles simulated with XFdtd. The HPol antenna gain is shown on the left, while the VPol is shown on the right. The full width of each dipole is 1.56~m, and they are elevated $\sim$3.96\,m above the ground over a $\sim$3.05\,m sign post. The antennas are simulated with a 200\,$\Omega$ characteristic impedance to model the 4:1~transformer. The simulated antenna sits in the center of a 150\,m ground flat plane tilted by 30$^{\circ}$ in elevation and $10^{\circ}$ from North to South. This configuration models the two antennas lower on the hill. An azimuthal angle (shown on the bottom) of $0^{\circ}$ corresponds to due East and an elevation angle (shown on the top) of $90^{\circ}$ looks directly up. The HPol beam pattern develops modes at a frequencies determined by the interference of ground reflections with the main lobe, while the VPol pattern appears to be mainly affected by the presence of the steel pole.}
    \label{fig:antenna-pattern}
\end{figure*}

Later modeling using antenna simulation packages NEC \cite{voors20154nec2} and XFdtd~\cite{xfdtd} suggested that the effect of the ground when looking near the horizon was too severe for a non-elevated antenna design.  Following this study it was determined that elevating the antennas off the ground was necessary to avoid ground interference; an antenna elevation height of $\sim$3.96~m was chosen as a compromise between performance and deployment difficulty. Ground effects are still present in the beam patterns, as shown in Figure~\ref{fig:antenna-pattern}; however, the ground planes used in the simulation are smooth planes rather than the rocky terrain seen at the site. Additional interference contributions are mitigated by avoiding any metal near the antennas in the support system.  

\begin{figure}[htbp]
\centering
\includegraphics[width=\iftoggle{double-column}{\columnwidth}{0.5\textwidth}]{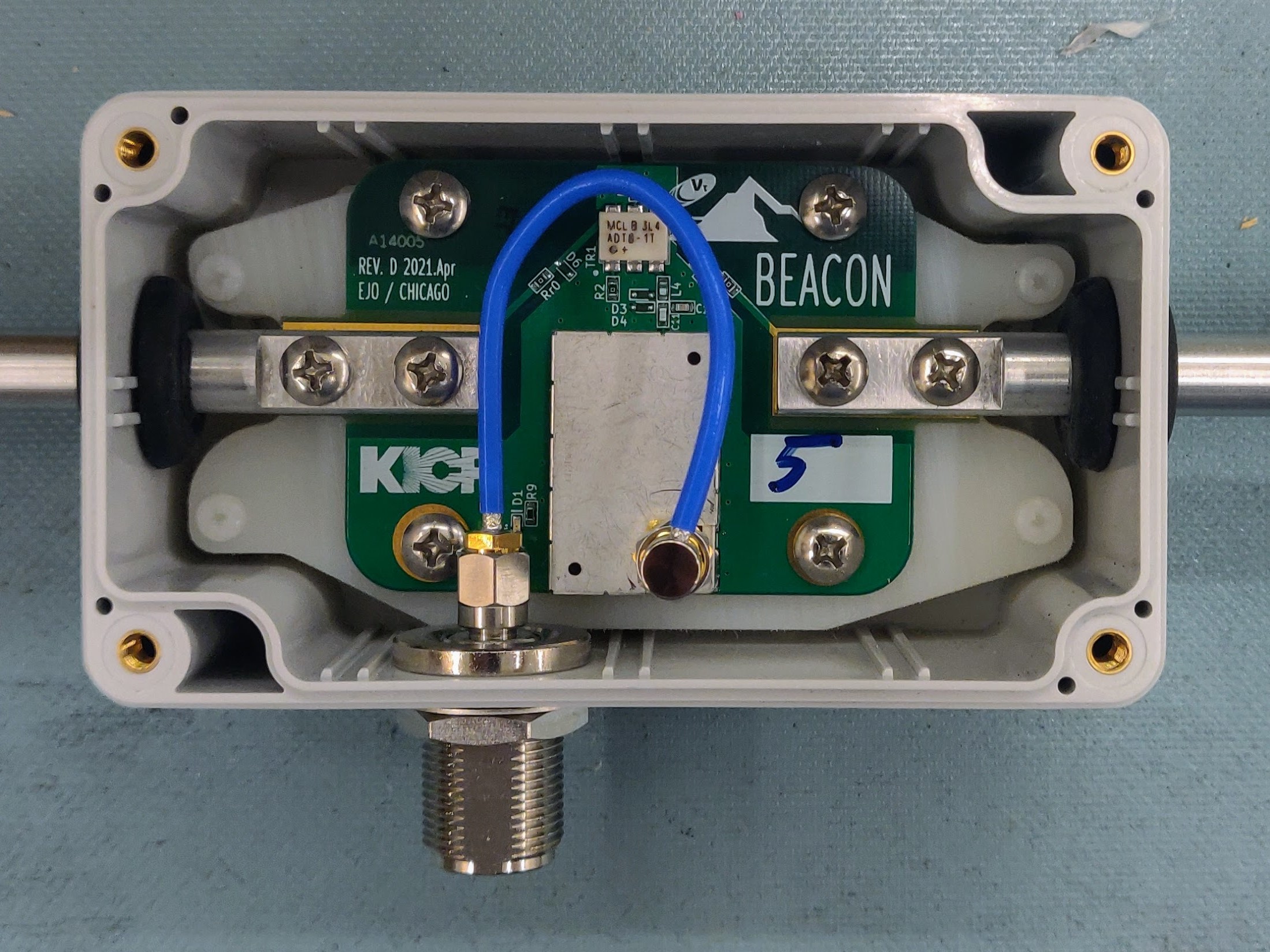}
\caption{Short dipole antenna feed.  The active dipole feeds each incorporate a 4:1 transformer into a 50~$\Omega$ LNA.  The Polycase enclosure helps protect the front-end board from weather.  The antenna elements are connected directly to the front-end board, with each extending outward through grommets in the enclosure.}
\label{fig:dipole_feed}
\end{figure}

\begin{figure*}[htbp]
    \centering
    \includegraphics[width=\textwidth]{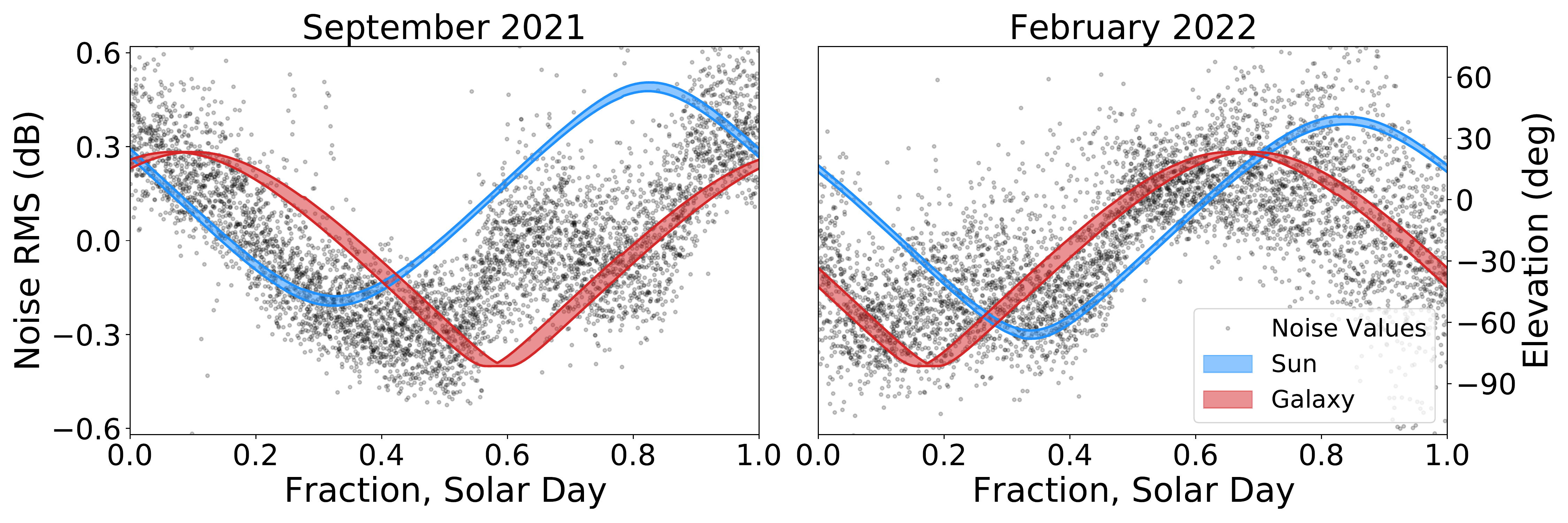}
    \caption{The fluctuations in the root-mean-squared (RMS) noise observed in VPol channel 5 during September 2021 (left) and February 2022 (right).  Superimposed on each plot are the range in elevation of the Sun and galactic core over the sampled time.  The RMS rises along with the galactic center, such that when the galaxy is visible in the antennas, the noise increases. The phase of the RMS variations follows that of the galactic center throughout the year, rather than the sun.
    }
    \label{fig:vpol-sky-noise}
\end{figure*}

We designed a custom short-dipole antenna with 2x76.2 cm (2x30 in) long tines with an active balun that could be mounted inside of a small enclosure on top of the mast, providing a low-profile and low-mass device capable of surviving the extreme environment. Although these antennas have a small effective height at the low-edge of the band, they provide a nearly omni-directional beam pattern across the band. Our BEACON active balun, shown in Figure~\ref{fig:dipole_feed}, consists of a 4:1 transformer that is fed into a 50~$\Omega$ low-noise amplifier (LNA), which is followed by a second stage of amplification. The transformer not only boosts the input impedance as seen by the antenna, but also isolates the common mode ground of the amplifier and coaxial cable from the dipole. To maintain a precision voltage to the on-board amplifiers, the balun is locally regulated to 3.0~V. The board draws \mbox{<\,45~mA} with a total gain of 35~dB.  Similar short dipole designs have been studied and utilized in the LOPES and CODALEMA experiments~\cite{Apel:2012uyt,Charrier:2015lsa}.

Figure~\ref{fig:vpol-sky-noise} shows that galactic noise is visible in the vertically polarized channel. Because of the small difference in period of the solar and sidereal days, we stacked the root-mean-squared (RMS) fluctuations in the noise over the course of a month at two periods of the year. When the galaxy rises above the horizon, the RMS noise is slightly elevated and the peak is correlated with the rising galactic center. We also note that this effect is not visible in the horizontal polarization, because the galactic center peaks in the South where there is a null in the HPol beam pattern. While the effect in VPol is weak, there is a clear phase shift correlated with the elevation of the galactic center at two different times of year. To be sensitive to this faint but pervasive galactic noise is a key goal of a transient radio detector like BEACON~\cite{Ellingson_radioarrays, 2001A&A...365..294D}.

\begin{figure}[htbp]
\centering
\includegraphics[width=\iftoggle{double-column}{\columnwidth}{0.5\textwidth}]{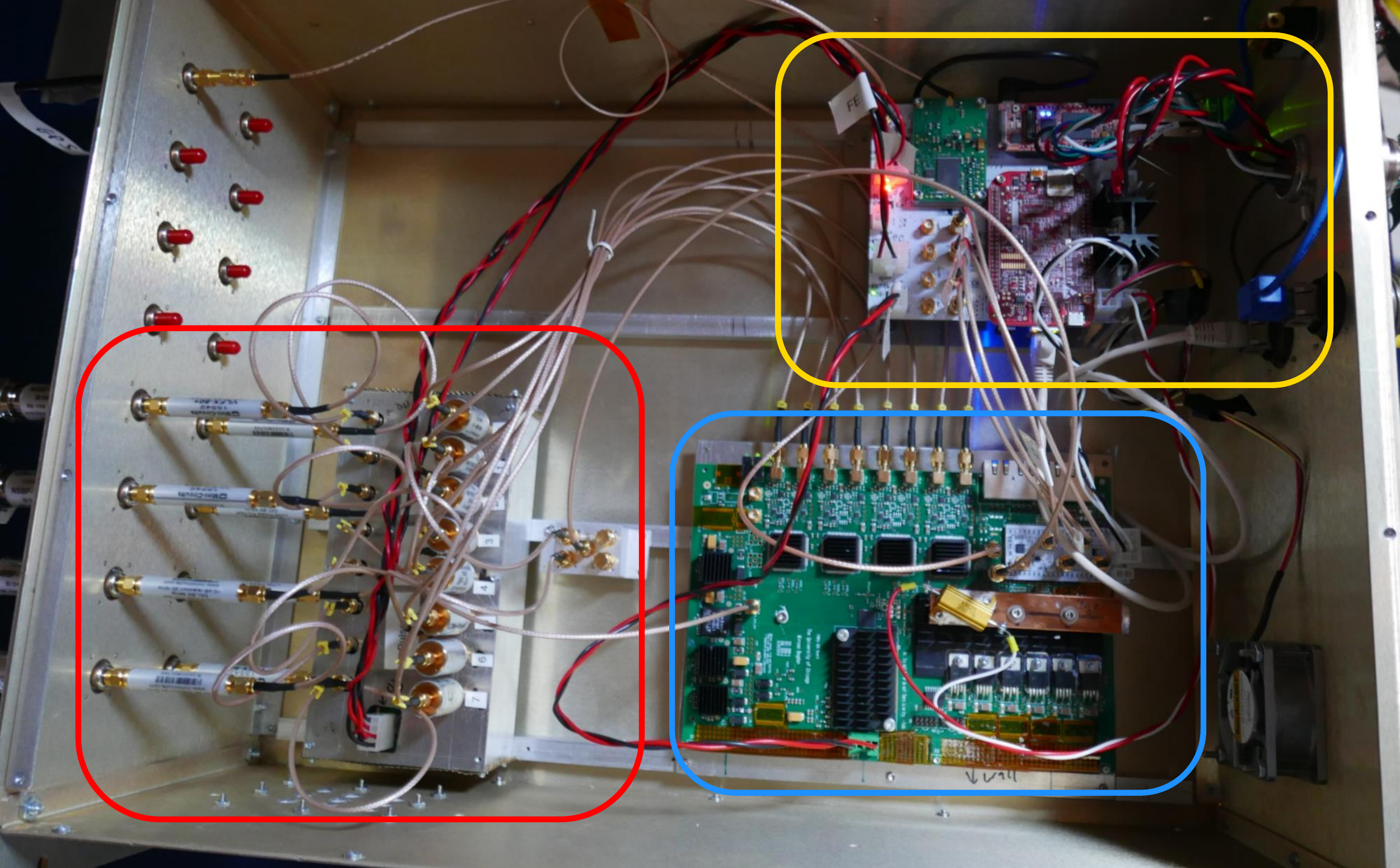}
\caption{Picture of the DAQ.  The yellow region in the top right contains the SBC, GPS clock, and power distribution.  The red region on the left contains second stage amplification and band-pass filtering.  The bottom right blue section is the custom digitizer and beamforming trigger board.}
\label{fig:DAQ}
\end{figure}

\begin{figure}
    \centering
    \includegraphics[width=\iftoggle{double-column}{\columnwidth}{0.5\textwidth}]{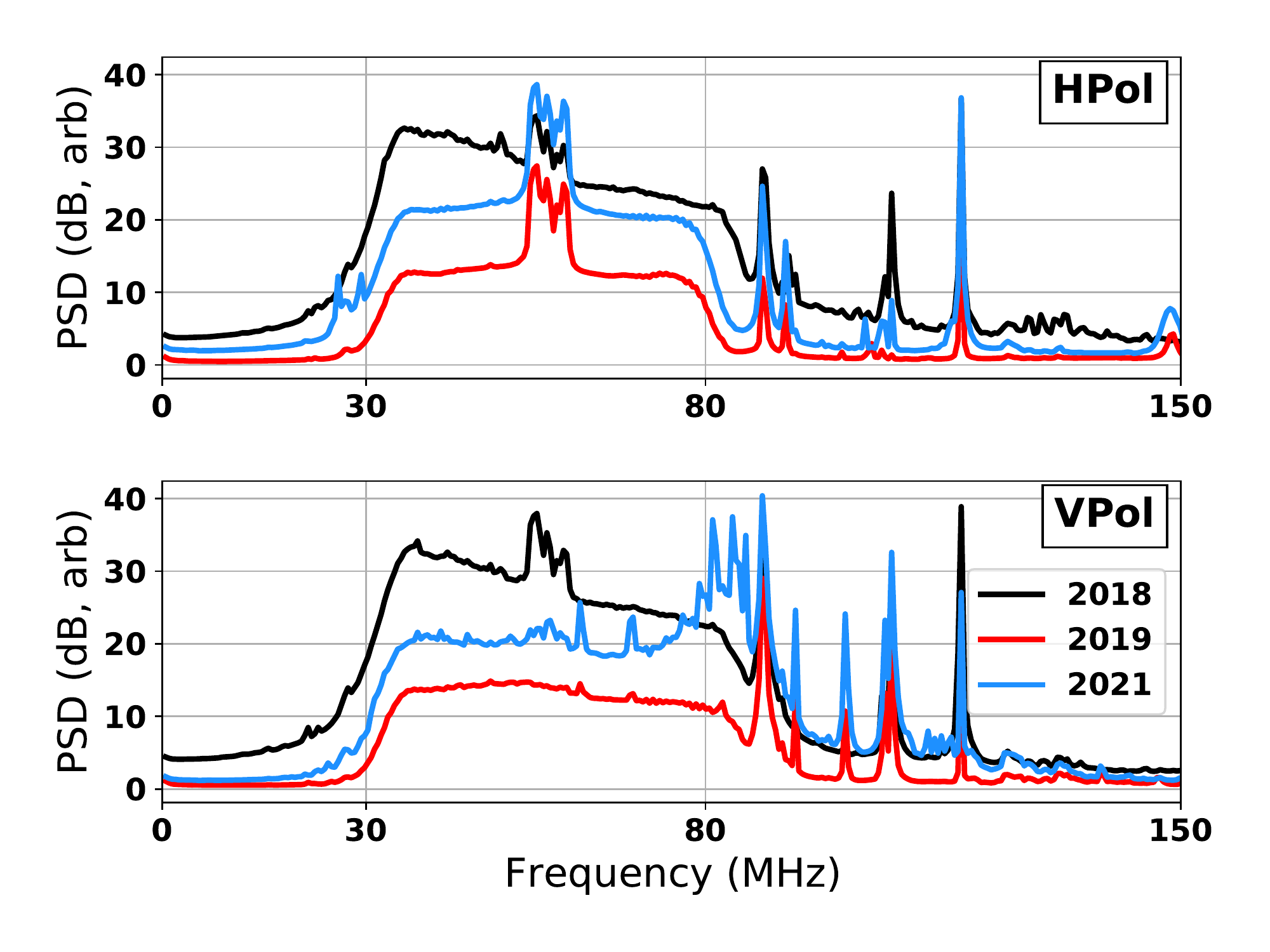}
    \caption{Top: Time averaged spectra for 3 generations of dipole antennas corresponding to the same HPol channel.  The time covered by each is set to be 50 runs, resulting in averages covering 391 hours for 2018, 133 hours for 2019, and 153 hours for 2021.  These times are sufficiently long for each generation that the differing time windows do not have a significant impact on the structure of the spectra. Bottom: The same except VPol antennas.  The spectra are presented as Power Spectral Density (PSD) in arbitrarily offset dB units (a conversion between ADU and volts has not been performed).  The variation in baseline power is a result of differing antenna construction and amplification which affects both signal and noise levels and is generally not representative of performance differences in SNR.  The 2018 traces correspond to LWA antennas, which were significantly lower to the ground and were generally a different infrastructure.  Comparing 2018 to other years it is clear that our VPol channel has significantly reduced cross-polarization power, as noticeable by the disappearance of television (TV) band noise in the VPol channel (with TV contributions ranging from $\sim$53~to 60~MHz, discussed further in Section~\ref{sec:rfi}).  The antenna element lengths were increased from 2x68.6~cm to 2x76.2~cm for the 2021 model, which has resulted in additional pickup in the high-end of the band, noticeable particularly in the VPol antenna which may be experiencing additional coupling with the steel mast due to closer proximity.  As the trigger operates primarily using HPol antennas this has not negatively impacted performance of the trigger.}
    \label{fig:antenna-comparison}
\end{figure}

\begin{figure}
    \centering
    \includegraphics[width=\iftoggle{double-column}{\columnwidth}{0.5\textwidth}]{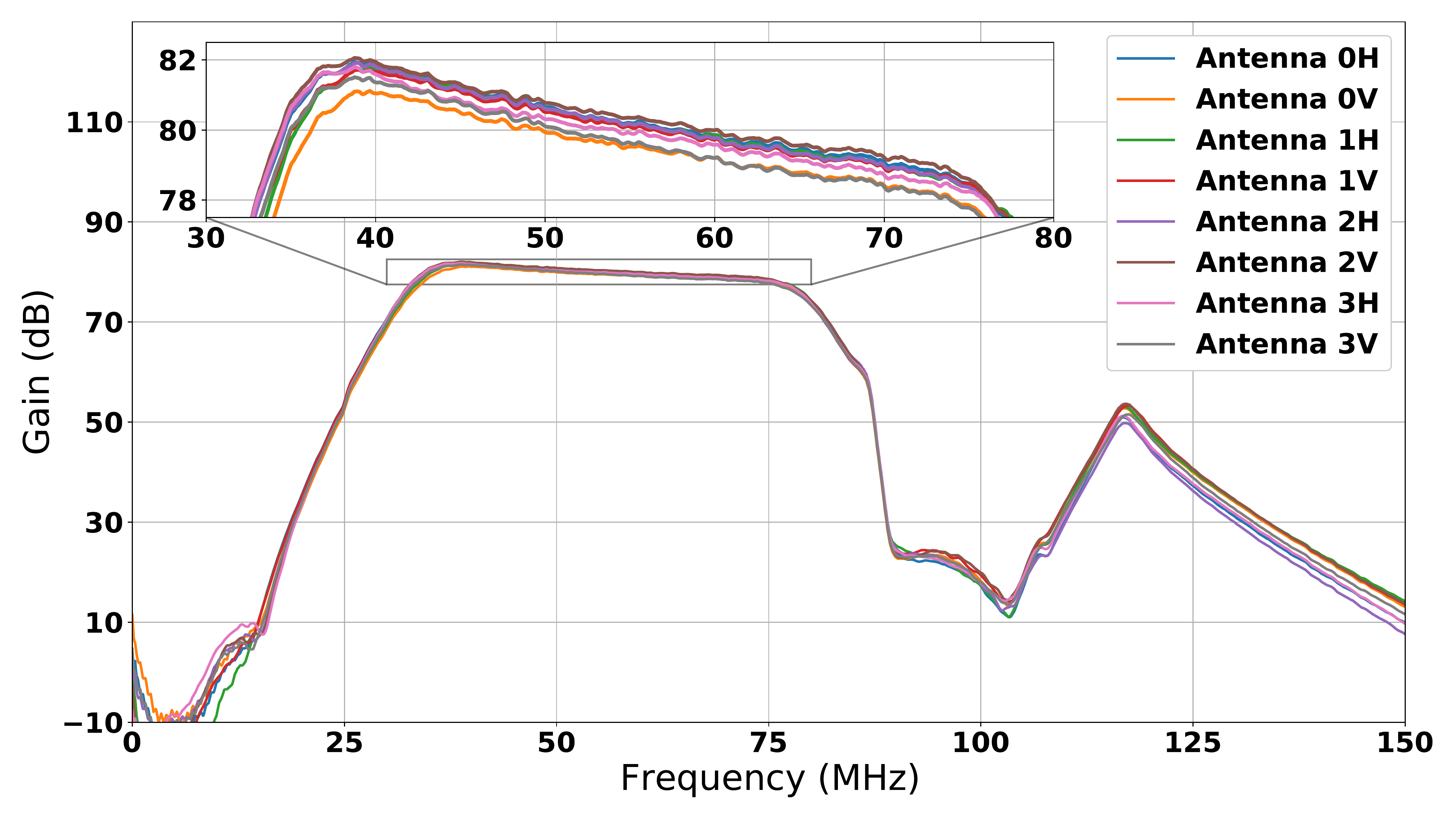}
    \caption{The system gain for each channel, including the active feeds with a gain of 45 dB,  cable losses through LMR400 and LMR240, bandpass and notch filters (Mini-circuits SHP-50, SLP-90, and NSBP-108), and second stage amplifier board with a gain of 40 dB. }
    \label{fig:response}
\end{figure}

The short dipoles are mounted directly onto a wooden masthead in a cross pattern for sensitivity in both horizontal and vertical polarizations; these are referred to as HPol and VPol antennas respectively.  A second HPol antenna could be a future addition for full angular sensitivity. However, the array's location on a mountainside reduces the need for sensitivity in directions parallel to the mountainside where the effective area is already significantly reduced.  The BEACON prototype is located on a North-South aligned ridge, so the HPol antennas are oriented North-South with the gain being maximized along the East-West axis, orthogonal to the ridge.  Because the Earth's magnetic field points close to the North, this orientation of the array aligns the center of the array's sensitivity with the direction that air shower radio emission is expected to be strongest (given by $\vec{v}\times \vec{B}$).

Coaxial cables ($\sim$107~m of LMR400 and $\sim$6~m of LMR240 in series) connect the antenna preamplifiers to the DAQ, carrying both the amplified signal and DC power to the preamplifiers, which have internal bias tees.  These cables are jumpered across the wooden masthead to the steel mast along with grounding cables, where they are guided to the ground before being run uphill towards the DAQ.  The cables are sheathed when on the ground to reduce damage from weathering as well as the local wildlife. 

Reliably elevating the antennas required a number of iterations, especially as the array location within Inyo National Forest precluded any permanent structures, requiring a mast support design that avoids drilling or pouring concrete. The first iteration of this design secured the base of the mast with a commercially available $\sim$33~kg rubber base, as well as 3 guy-lines tied to local rocks for each mast.  This design was improved in follow-up deployments in 2020 and 2021, which addressed issues with failed wooden mastheads and fallen masts.  These issues were caused by the extreme weather at the prototype site, with gusts of up to 130~km/h, heavy snow and ice build-up, static discharges, lightning strikes, and exposure to the sun.  The improved supports included $\geq$ 6 guy-lines per mast using higher-grade rope, and 4 wooden struts per mast.  The struts are cut to length on-site such that they can be wedged securely into the local terrain (Figure~\ref{fig:antennas}). 

This upgraded design is robust to animals climbing or pulling on it, is readily adaptable to varied terrain and has proven to be capable of withstanding winter conditions. In places with fewer restrictions, drilling into the ground would add additional stability. Though the wood used was high-quality pressure-treated cedar, it still showed significant weathering after just a single year, so improvements in RF-safe alternative materials to wood, such as fiberglass, for the masthead may be warranted for future deployment, while steel struts could be an option for usage away from the antennas.

\subsection{Radio Frequency (RF) Signal Chain and Data Acquisition System (DAQ)\label{sec:daq}}

The BEACON prototype uses a custom DAQ housed in a Faraday enclosure, shown in Figure~\ref{fig:DAQ}. At the input, signals pass through a lightning arrester bank to prevent static discharge from damaging the system. Afterwards, signals then pass through an RF receiver board, which provides 35 dB of amplification, a DC bias for the antenna feeds, filtering, and power limiting. Filters include both 30-80 MHz band-definition filters as well as FM notch filters, which are necessary due to the proximity of the FM broadcasts. Typical noise spectra for three generations of BEACON antennas are shown in Figure~\ref{fig:antenna-comparison}. Figure~\ref{fig:response} shows the combined gain of the full RF signal chain. 

Signals then reach the  digitizer and beamforming trigger board, which incorporates 8 channels of 500 MSPS 7-bit digitization and a control FPGA responsible for triggering and buffering up to 2048 samples per channel for readout once triggered.  Typically, only 1024 samples are read out per event to increase readout speed and reduce dead time and data volume. Tunable attenuators allow for gain matching between channels and tuning the dynamic range of the digitizer. A timing GPS is used to provide a reliable pulse per second (PPS) to the digitizer board, which records the number of clock cycles when the PPS is received, allowing for precise time tagging for each trigger. 

The digitizer and beamforming trigger board is controlled and read out via SPI using a BeagleBone Black (BBB) single board computer (SBC), running Debian Linux. Software on the SBC manages configuration of the trigger, readout of event data and metadata, housekeeping, and transfer of the data off of the DAQ. The BBB is connected to the Barcroft network via Ethernet. Also on the network is a microcontroller which allows for remote power cycles of the entire system.

The DAQ system is powered by a 15~V DC supply, plugged into the Observatory Dome power system. The present power draw of the DAQ is $\sim$40 W, dominated by the digitizer and trigger board. Power at the Observatory Dome is provided by a solar-battery hybrid system deployed by WMRC.  The typical power system capacity is considerably greater than the daily usage ($\sim$1~kWh), resulting in nearly complete live time, except under extended extreme cloud cover or excessive snow lasting \mbox{> 5 days.} 

Data is sent from the DAQ system to our Archive Machine computer located nearby at Barcroft Field Station. The Archive Machine archives data before it is transferred to the University of Chicago and provides local monitoring. In addition to being connected to the Barcroft network, the Archive Machine is also connected to a backup cellular network, which allows communication with the DAQ system when the normal connection from Barcroft is down, a relatively common occurrence particularly in winter. Also at Barcroft is a Raspberry Pi with a software-defined radio tuned to listen to aircraft ADS-B transmissions, which is used alongside data provided by The OpenSky Network~\cite{opensky} for the purpose of tracking nearby commercial airplanes.  The use of ADS-B transmissions to correlate airplane locations with above-horizon RF signals in the 30-80 MHz band has been demonstrated previously by other experiments~\cite{PierreAuger:2015aqe}.

\subsection{Trigger System\label{sec:trigger}}

The combination of an FPGA and streaming digitizer on the digitization and trigger board allows for flexible triggering capabilities. Currently, a beamforming (i.e. phased array) trigger is implemented, similar to the one deployed as part of the Askaryan Radio Array (ARA) at the South Pole~\cite{allison2019design}.  This trigger uses a pre-calculated table of expected arrival time differences between the antennas to delay signals before summing them.  Each set of time delays corresponds to a beam sensitive to a particular direction, and is most sensitive to signals arriving from the specific direction where the delays result in coherently summed signals.  This coherent sum will increase an incoming signal's voltage by a factor of $N_\mathrm{antenna}$, while thermal noise will add incoherently and only increase as $\sqrt{N_\mathrm{antenna}}$, resulting in a net SNR increase of $\sqrt{N_\mathrm{antenna}}$~\cite{Vieregg_2016}.

The delayed and summed waveforms are further processed in the DAQ with a ``power sum''.  This is done by first squaring the combined signal to obtain a proxy for power, before summing the combined power signal in 16 sample (32\,ns) bins every 8 samples (16\,ns) such that each bin has some overlap with the previous bin.  The power sum of a coherently summed signal will increase the power SNR by a factor of $N_\mathrm{antenna}$. Hereafter we refer to the beam SNR calculated by aligning voltage waveforms and summing as the ``beam voltage SNR'' and the power sum performed in the triggering hardware as the ``beam power SNR''. Currently, the time delays for each beam are pre-calculated assuming plane wave signals, however near-field time delays could be implemented in the FPGA if desired.

The trigger rates in each beam are continuously monitored and the thresholds are adjusted to meet user-defined goals. In this way, trigger thresholds are dynamically set to be noise riding, managed by the SBC such that a global trigger rate of 10 Hz is maintained.  With a target rate of 10 Hz and temporary system outages accounted for, we conservatively estimate our dead time to be $\sim$1.5-2\% over the span of time used in the analysis discussed in Section~\ref{sec:analysis}. The thresholds for each beam are adjusted automatically and in nearly real time so beams with consistently loud sources of RFI do not dominate the trigger. The rates of the individual beams can be further refined by the user. This directional trigger is essential in RFI-rich environments and has allowed the BEACON prototype to maintain relatively low thresholds in the majority of beams despite prevalent RFI from certain directions. 

\begin{figure*}[tbp]
\centering
\includegraphics[width=\textwidth]{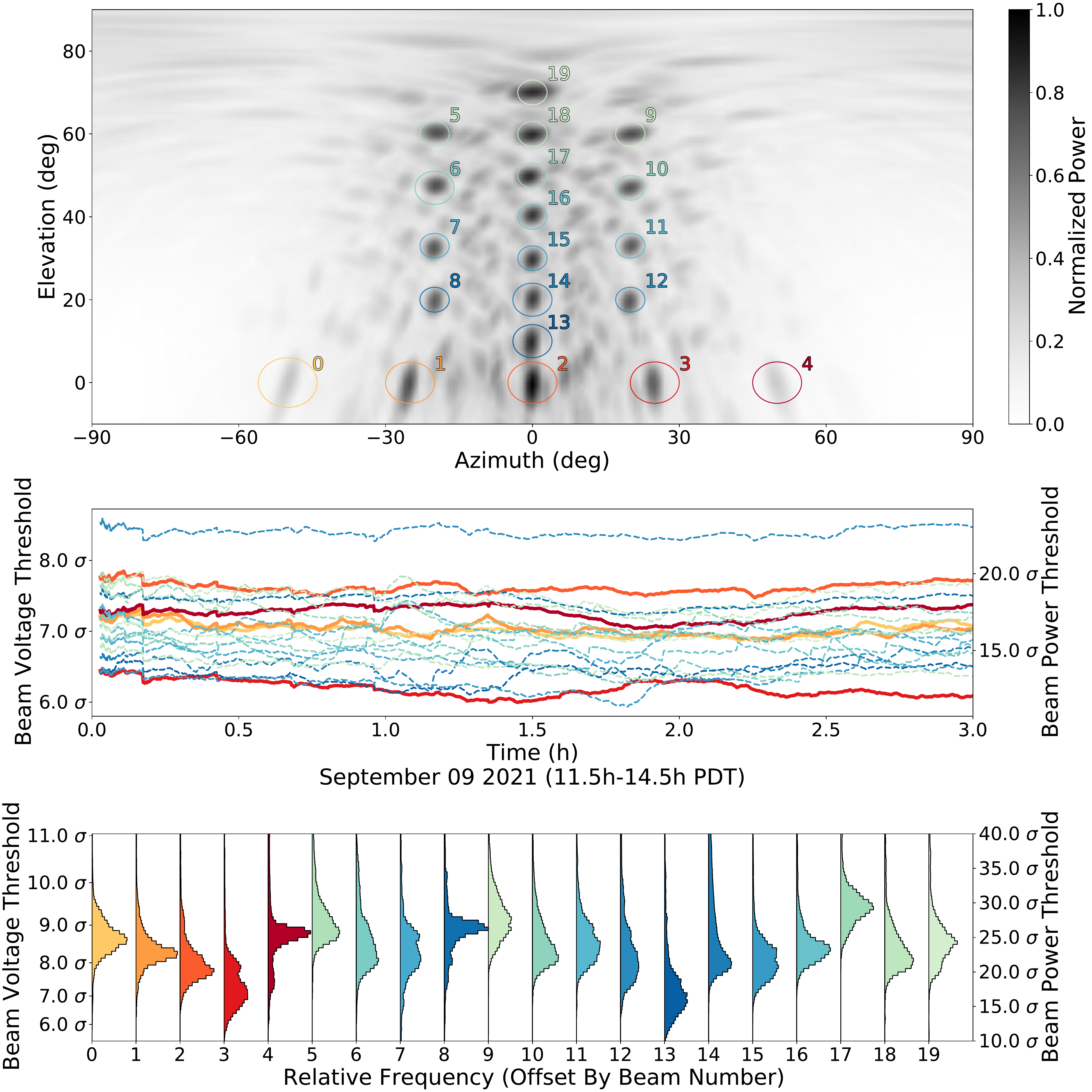}
\caption{Top:  The current beam map, with gray-scale color map corresponding to the normalized maximum power perceived in any beam for a mock signal arriving from each point on the map; maximal sensitivity/power is achieved in the nominal directions of each beam.  Each beam is labeled and circled with radius set to 3~dB below that beam's max power.  Middle:  The thresholds for each beam during a quieter run. The measured power SNR, referenced to the instantaneous noise from the DAQ, is shown on the right axis. Beam voltage SNR is shown on the left axis and is computed from cosmic ray simulations as described in the text. Colors of each line correspond to the same colors used in the top plot, with beams near the horizontal being solid red lines, and above horizontal beams being dashed blue and green lines.  The near-horizon beams generally exhibit a higher power threshold, as expected from anthropogenic noise.  Some above-horizon beams point to prominent sidelobes of below-horizon RFI, and will also show elevated thresholds. Bottom: The long-term distribution of thresholds in each beam over the $\sim$112 day period discussed in Section~\ref{sec:analysis}.}
\label{fig:thresholds}
\end{figure*}

Currently we use 20 beams, distributed as shown in the upper panel of Figure~\ref{fig:thresholds}.  These beams were optimized for triggering on above-horizontal events in the region expected to be populated by cosmic rays.  A full-scale BEACON station would be targeting the near-horizon region where tau neutrino events are expected. Moreover, the total number of beams would be expanded to uniformly fill the aperture. The trigger logic for the original implementation with ARA is described in more detail in Reference~\cite{allison2019design}.

The trigger implementation allows for additional calculations to be performed in order to form noise-rejection vetoes to improve performance.  Some examples of vetoes that have been considered: \begin{enumerate}
    \item a ``side-swipe'' veto, which can actively veto events where the amplitude on one antenna is significantly larger than others;
    \item a saturation veto for high-power events which are clipped significantly; 
    \item a rear-facing veto which would avoid triggering when an event hits both western antennas first;
    \item and a ``band ratio'' veto which compares relative power seen through 2 finite impulse response (FIR) filters in the low and high portions of the band to reduce triggers from narrow band events.
\end{enumerate} 

Beams pointed at specific known sources of RFI can also be used as a veto. As shown later in Figure~\ref{fig:resolution}, regular anthropogenic sources can be localized to well within the beam width. A veto could be implemented that disallows events from a certain beam direction if it also triggers a sideband.

The performance of the phased trigger can be seen in Figure~\ref{fig:thresholds}, which illustrates each beam's definition and dynamic thresholds.  Thresholds are computed as the power sum over a 16 sample (32\,ns) window in each beam and are shown here referenced to the RMS noise in a beam, which is monitored continuously by the DAQ. In the middle and bottom panels of Figure~\ref{fig:thresholds}, the power SNR thresholds are compared to the voltage SNR thresholds in the beams. The translation between the power thresholds used in the triggering hardware and the voltage thresholds shown on the left are computed from simulations of cosmic rays modeled with ZHAireS and propagated through the prototype signal chain~\cite{Zeolla:2021bl}. The translation is $V_\mathrm{SNR} = 1.8 \sqrt{P_\mathrm{SNR}} -0.38$. We make this comparison because prior simulation studies used beam voltage SNR thresholds for modeling the tau neutrino sensitivity~\cite{wissel2020prospects}, while the trigger hardware uses thresholds on the beam power SNR.  The thresholds achieved on the instrument approach the nominal thresholds assumed in the simulation studies (5$\sigma$ in voltage)~\cite{wissel2020prospects}. While the thresholds are often in the range assumed by the simulations, there are also periods of time dominated by loud RFI in the field of view.

Comparing the beam map to the thresholds, we can see that there is some variation in the thresholds in each beam, corresponding to the observed rates in those beams. Some beams, like beams 0 and 4, point near a source of RFI below the horizon and maintain a higher mean threshold compared to other beams at the same elevation. Other beams, like beams 5, 9, and 17, point well above the horizon but may be triggering on sidelobes from RFI below the horizon.  The impact of sidelobes is expected to be reduced with the increased number of antennas in a full-scale BEACON station trigger array.

\section{Instrument Performance and Data Analysis\label{sec:analysis}}

\subsection{Antenna Position Calibration and Direction Reconstruction\label{sec:calib}}

BEACON uses interferometry of waveforms from each of its 8 channels to determine the source direction of signals.  Accurate source direction reconstruction (both at the trigger level and in analysis) requires precise knowledge of the array timing, including the location of each antenna and signal cable lengths.  A calibrated array can use pointing for RFI rejection of permanent sources or airplanes, as well as for characterization of the polarization and source properties of the initiating radio source.  The typical method of calibrating an array consists of the following process:
\begin{enumerate}
    \item Perform initial position measurements using a Global Positioning System (GPS) system. 
    \item Measure cable delays using a vector network analyzer or time-domain reflectometer (TDR)
    \item Send radio pulses at the array from a known location (also measured with comparable system to antennas), recording the pulses through the DAQ for later analysis
    \item Perform a $\chi^2$ minimization targeted at matching predicted arrival time differences in each channel with the actual measured delays through the DAQ for various pulser locations or sources.  A typical $\chi^2$ takes the form:
    \begin{equation}
        \chi^2 = \sum_j^{n_\mathrm{sources}} \sum_i^{n_\mathrm{baselines}} \frac{\left[t_\mathrm{geometry, i,j} - t_\mathrm{measured, i,j}\right]^2}{\left[\sigma\left(t_\mathrm{measured, i,j}\right)\right]^2}\label{eqn:phase_center_fit}
    \end{equation}
    where the antenna positions (geometry of the array) are adjusted at each iteration until minimization has been obtained.  
\end{enumerate}

The number of degrees of freedom (DoF) for the minimization parameters is equal to 3 axes of movement + 1 cable delay per antenna, leading to 16 total DoF.  The number of distinct measurements provided by $N_\mathrm{site}$ pulsing sites is given by $\mathrm{DoF} = N_\mathrm{site} \cdot C( N_\mathrm{antenna} , 2 )$ where $C( n , r )$ is the choose operator, which determines the number of arrival time differences (baselines) that can be calculated when comparing 2 antennas from a set of $N_\mathrm{antenna}$.  Minimization was performed independently for each polarization, allowing for variations in phase centers between HPol and VPol antennas.  Pulsing locations are chosen to be far from the array such that variations in timing from uncertainties in their locations are negligible and do not add additional DoF to the minimization.

Initial position measurements of the antenna masts were made with the Real Time Kinematic (RTK) technique that compares GPS positioning of two nearby GPS antennas - resulting in cm level precision by correcting for the propagation conditions in the local atmosphere~\cite{rtk}. Each antenna mast includes a dual-band GPS patch antenna, which may be connected to a GPS receiver on demand. We use a u-blox C099-F9P application board (ZED-F9P GPS~\cite{zed-f9p-manual}) to measure the position of each antenna, with corrections provided by a UNAVCO GPS station permanently installed $\sim$30~m away from the BEACON site at 37.58915N, 118.23844W~\cite{unavco}. 

Pulsing data was taken over the course of 3 days during a calibration campaign in 2021, during which pulsing data was taken for 6 separate sites in both polarizations.  The transmitter included a high-voltage pulser (FID technologies FPM 10-1PNP) driving a biconical antenna (Aaronia BicoLOG 30100E) at known rates, with varying fixed attenuators. These pulsing data are used in the fit described in Equation~\ref{eqn:phase_center_fit}.  The resulting errors on the phase centers are estimated to be less than 5\% of the shortest relevant wavelength.

The source direction is reconstructed using interferometry~\cite{Romero-Wolf:2014pua}. Cross correlations are calculated for each pair of antennas in separate polarizations. The cross correlation for a given antenna pair is expected to peak at a time delay consistent with the arrival direction of the signal. A ``correlation map'' is formed by sampling these cross correlations at delays expected for each direction. Each direction in the map corresponds to the average correlation value from each baseline when sampled at the expected time delay for that direction.  The expected time delays depend on the geometry of the array and source direction on the map and are often calculated assuming a plane wave for distant sources. The peak value of the map corresponds to the direction which has sampled each baseline's cross correlation nearest the maximum.  Each baseline is weighted equally and is normalized such that identical signals result in a maximum cross-correlation value of 1; a map generated with identical signals in each channel would also result in a peak value of 1.  Real signals vary slightly across antennas, so the optimal map value depends on each event and is typically \mbox{< 1}.  

A perfect impulse would have a single peak in a cross correlation, resulting in a single ring of possible arrival directions on the sky for each baseline due to the symmetry around the axis connecting those 2 antennas.  By averaging maps of all 6 baselines, the degeneracy of these rings is broken, with all baselines overlapping only in a single location for an impulsive plane-wave in a properly-calibrated array.  This requires a sufficient number of baselines to fully break degeneracy, or ambiguities in pointing can occur.  Though the BEACON prototype has a sufficient number of antennas to accurately point to most impulsive RF sources, narrow-band signals result in highly periodic cross correlations which in turn produce a series of concentric rings on the maps per baseline, increasing the degeneracy of potential source directions.  This problem can be exacerbated by the presence of unrelated continuous wave (CW) noise, coincident signals from other RFI, or by the source signal itself being insufficiently impulsive. 

After minimization, signals from the mountainside pulsers show maximal reconstruction offsets of $\sim$1$^\circ$, with the majority of sites showing offsets $<$0.5$^\circ$.  The accuracy is discussed further in Section~\ref{sec:rfi} when presenting airplane reconstructions, which provide an external source of signals with known directions and show a systematic offset of ~1-2$^\circ$.  The mountainside pulsers provide a limited range of elevation angles for the calibration minimization, which could contribute to the observed reconstruction error.  Additionally, the cable delays can have a degenerative effect with antenna position within the minimization for adjusting baseline timings, which also could be the source of the discrepancy.  In future efforts we aim to address these issues with a drone pulser (Section~\ref{sec:conclusion}), which would provide a significant increase in angular range used for calibration. 

\begin{figure*}[htbp]
    \centering
    \includegraphics[width=\textwidth]{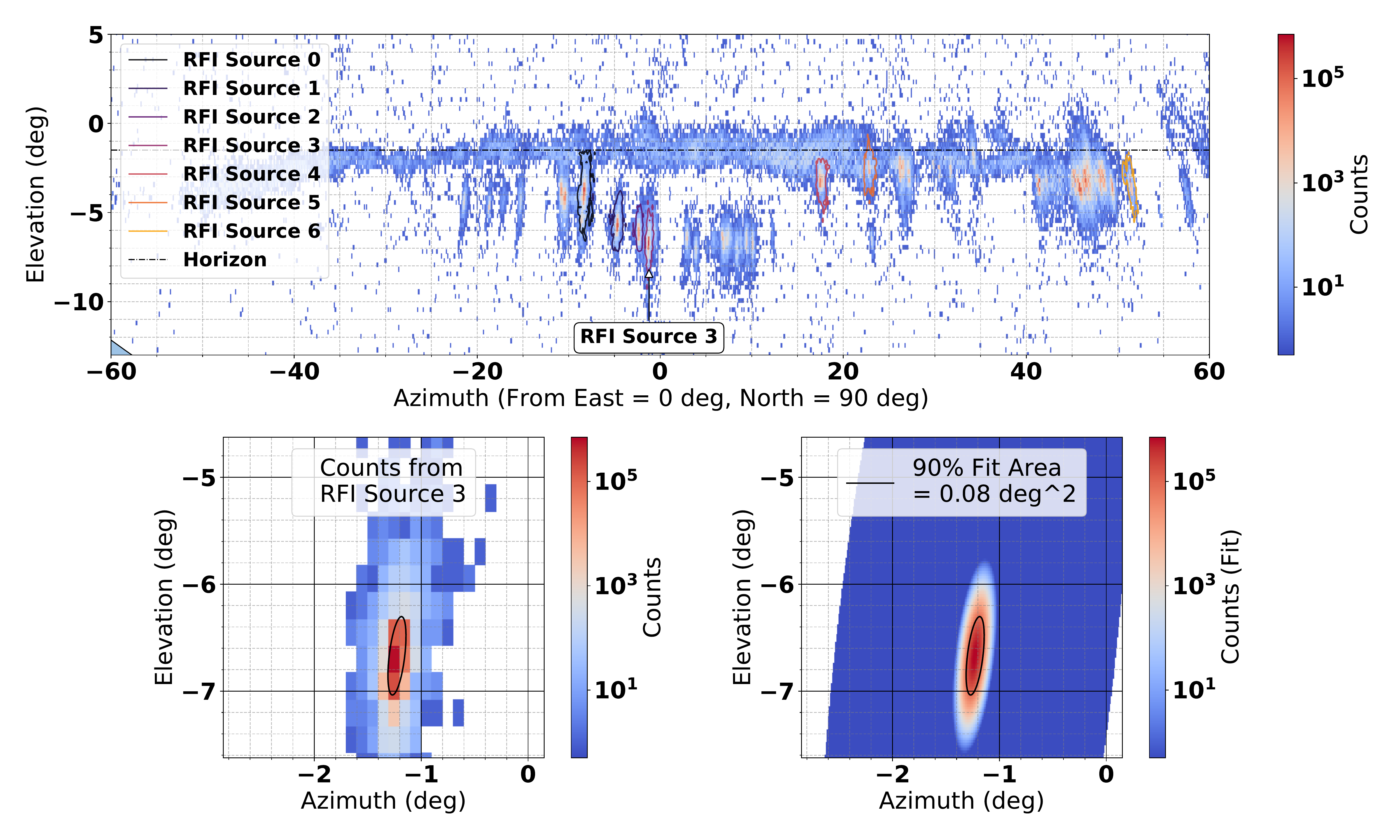}
    \caption{Top: Reconstruction direction of events from one week in September 2021.  Seven of the most populated RFI sources have been highlighted.  These sources are fit with a two-dimensional Gaussian after isolating the events in each region. Bottom Left: Isolated events from RFI Source 3 (arbitrarily chosen as an example).  Bottom Right:  2D Gaussian fit (color map), with outline of the 90\% integral area of the fit plotted on top.  Note that the color scale is logarithmic and represents counts for all 3 plots.  The average fit 90\% integral area for all 7 sources was \mbox{<\,0.1} sq.\@ degrees.  The approximate location of the horizon has been indicated at an elevation angle of -1.5$^\circ$.}
    \label{fig:resolution}
\end{figure*}

The precision of the prototype to reconstructing the arrival direction of stationary radio signals was experimentally determined by reconstructing arrival directions of below-horizon RFI sources, the majority of which arrive from a few very loud stationary emitters.  A 2D Gaussian was fit to 7 of the most prevalent sources, with an average 90\% integral area for the fits of \mbox{<\,0.1} sq.\@ degrees (see Figure~\ref{fig:resolution}).

\begin{figure}[htbp]
\centering
\includegraphics[width=\iftoggle{double-column}{0.8\columnwidth}{0.4\textwidth}]{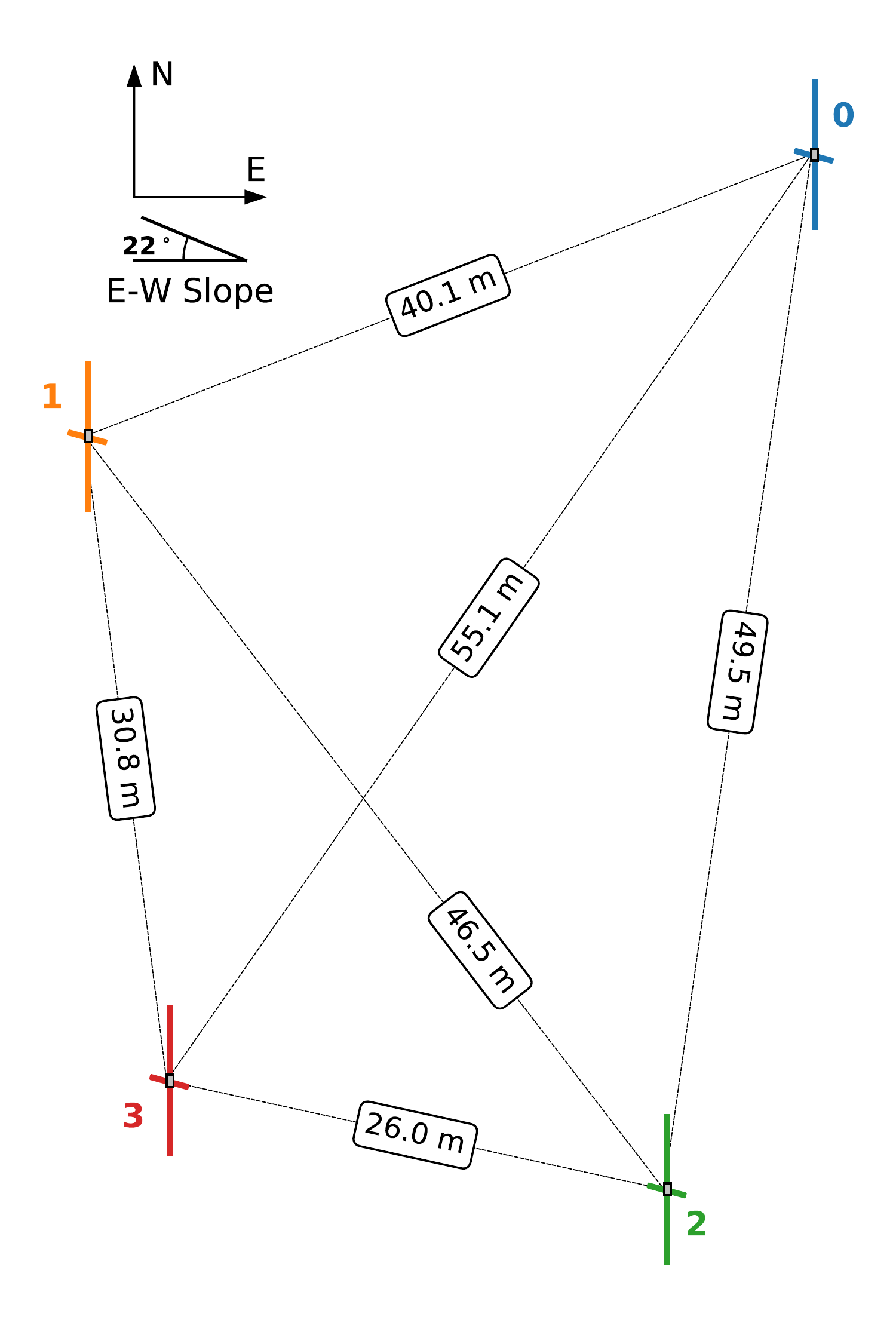}
\caption{A top-down view of the array layout in local East-North-Up (ENU) coordinates.  Positions correspond to calibrated HPol phase centers.  Baseline distances have been labeled for each antenna pair.  The slope the antennas are situated on is rugged and generally amorphous, however the approximate downhill slope across the array in the East-West direction is $22^\circ$.  Relative to the lowest antenna (mast 0), the heights of 1, 2, and 3 are approximately \mbox{15.9 m}, \mbox{4.0 m}, and \mbox{13.7 m} respectively.  The size of each antenna has been magnified 5$\times$ compared to baselines for visibility.}
\label{fig:position-schematic}
\end{figure}

\subsection{Characterization of RFI\label{sec:rfi}}

The BEACON prototype instrument is positioned near the border of California and Nevada and looks East over the Nevada desert.  This region is populated by a series of small towns with agricultural and mining industries, military bases, and power infrastructure like the Crescent Dunes Solar Project.  The site is also just south of a common commercial air flight path.  Though all of these anthropogenic sources are tens to hundreds of kilometers away from the site, many sources are visible to BEACON due to its high elevation and sensitivity to radio signals at signal strengths near thermal levels.  

In this Section, we discuss several ways this anthropogenic activity appears in the data taken with the prototype instrument.  The vast majority of anthropogenic signals in the data can be easily separated from cosmic-ray and neutrino signals due to signal shape, polarization, spatial and temporal clustering, and other event characteristics.  

 \textit{Static Sources}: The most common category of events come from towns and infrastructure.  These events cluster spatially, are expected to be localized to a single beam, and therefore can be cut based on their direction.  The signal shapes observed from different RFI sources can vary significantly, however signals from a single source are generally very consistent. 

\begin{figure}[htbp]
\centering
\includegraphics[width=\iftoggle{double-column}{\columnwidth}{0.75\textwidth}]{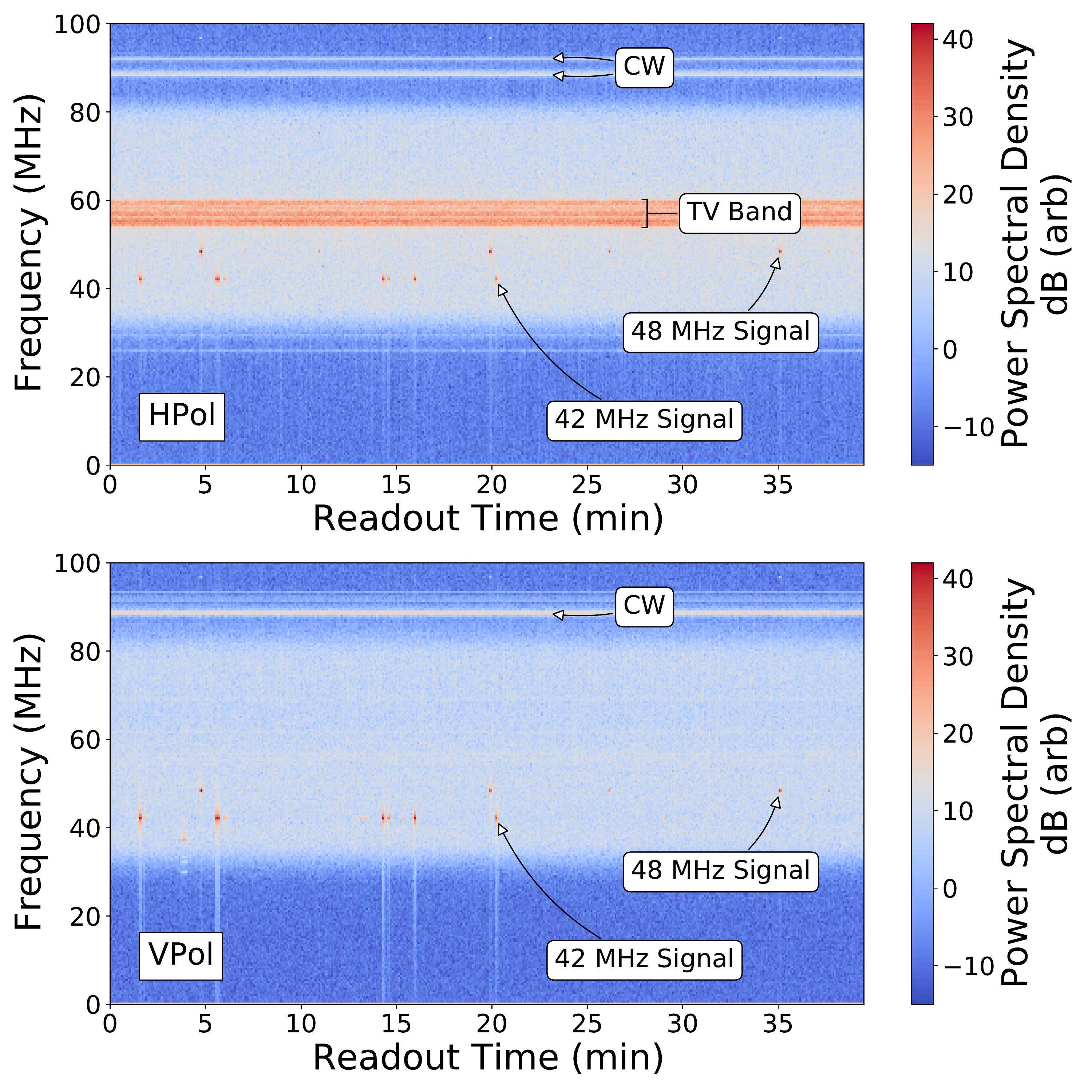}
\caption{Spectrograms of the HPol (top) and VPol (bottom) channels of antenna 0 generated using untriggered events recorded once per second during a run in October 2021.  Several features are highlighted in the spectrograms, including examples of CW noise, the TV broadcasting band, and intermittent RFI at 42 and 48 MHz associated with radio communications.  The color map is presented in arbitrarily offset dB units (a conversion between ADU and volts has not been performed).  Strong CW sources are capable of saturating the system, which results in aliasing that is apparent as a coincident broadband increase in power within the spectrogram.}
\label{fig:waterfall}
\end{figure}

 \textit{Continuous Wave Sources and the Television Band}: CW signals are narrow band, arriving at the array with very little temporal variation.  Because of this they are often not directly responsible for activating the trigger (which is designed for temporally impulsive signals); however, they are commonly visible in the spectrum of triggered broadband signals.  We typically remove these from the data in offline analysis via notch filters and the sine subtraction technique described in Section~\ref{sec:abovehorizon}.  The nominal band of the BEACON antennas overlaps with common radio communication frequencies, as well as the low-VHF television (TV) broadcasting range.  Signals from the KHSV TV station in Las Vegas are pervasive in all HPol data, despite the transmitter being over 300 km away and lacking a direct line of sight to BEACON.  A notch filter is currently used in analysis exclusively in HPol channels to combat this signal.  Figure~\ref{fig:waterfall} shows the spectra of HPol and VPol antennas over the course of a few hours.  Bright horizontal lines in these plots correspond to CW sources.  The TV band is visible in HPol from $\sim$53~to 60~MHz.  Intermittent short bursts of activity can be seen at 42~and 48~MHz, which are associated with communication systems for the California Highway Patrol and the Los Angeles Department of Water and Power, respectively.

 \textit{Periodic Noise Sources}:  An excess of events have been observed to arrive at the BEACON prototype with time differences corresponding to multiples of  \mbox{1/(60 Hz)}.  These signals can be associated with arcing or similar discharges from power infrastructure, which operates at 60 Hz in the US. A 55 kV high-voltage transmission line connecting Nevada with the Owens Valley runs within the field-of-view of the prototype, with several substations.  When there is snow on the valley floor, this class of signal largely is suppressed, perhaps because the snow is acting like an insulator to prevent arcing.
    
\begin{figure*}[htbp]
\centering
\includegraphics[width=\textwidth]{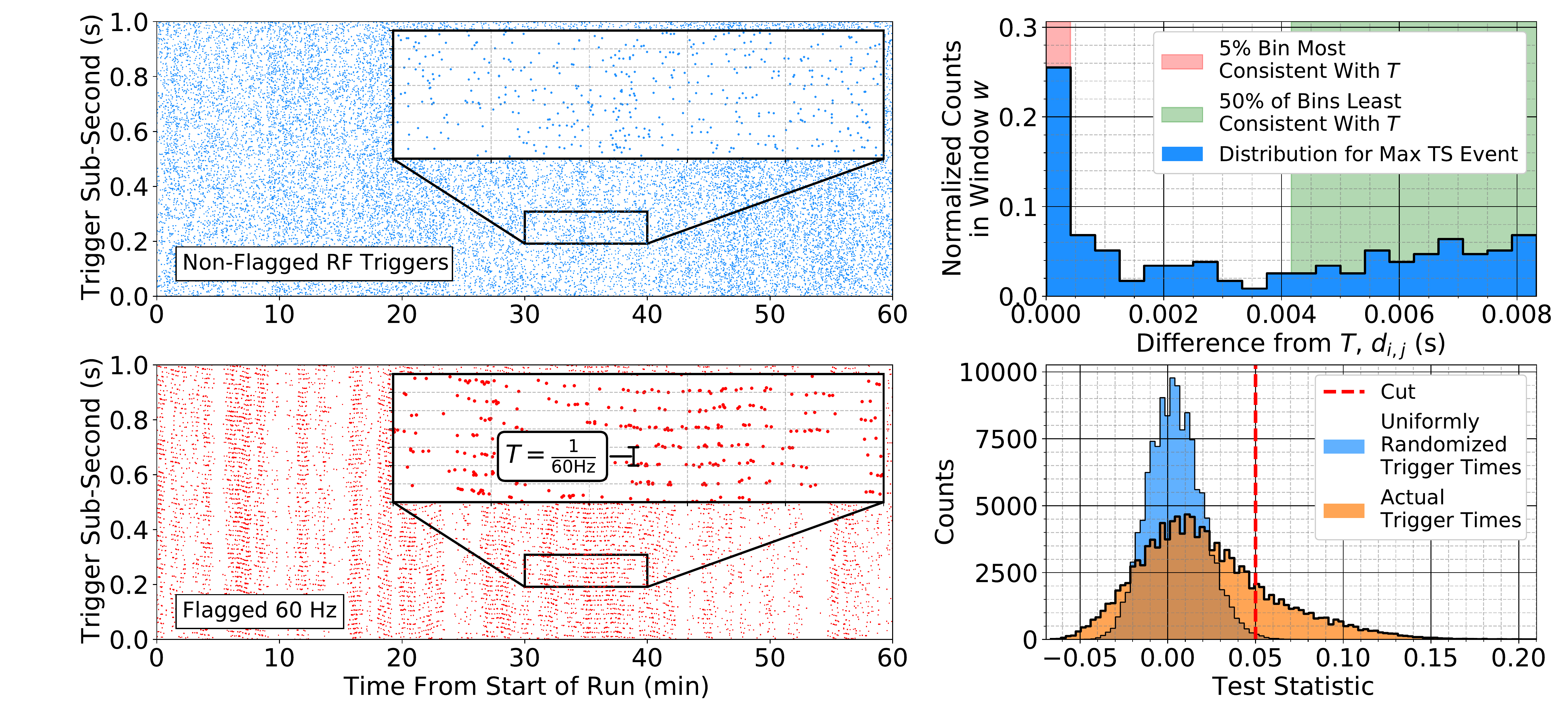}
\caption{Left: The arrival time of RF-triggered events within a run from September 2021, with sub-second timing plotted on the y-axis; events not flagged are shown in the top left, with events flagged by the algorithm to be consistent with an arrival rate of 60 Hz (with corresponding periodicity of $T = 1/60$~s) shown in the bottom left.  Insets show striations in the bottom plot consistent with the expected periodicity.  These flagged events represent $\sim$20\% of the total events in the 1 hour span shown.  Top Right: Histograms showing the portion of events arriving at an interval consistent with $T$ for the highest test statistic ($\mathrm{TS}$) event.  The $\mathrm{TS}$ is the difference in counts in the red region to the mean of the green region.  Histograms created using window \mbox{$w = $ 20 s.}  Bottom Right: A histogram of all $\mathrm{TS}$ values for this run.  An example cut has been applied near the limit of the $\mathrm{TS}$ as calculated for uniformly distributed trigger times, beyond which events are highly likely to be consistent with $T$.  The events flagged will be used to motivate further targeted cuts based on direction, template matching, and signal properties, to further improve the efficiency for removing this form of RFI.}
\label{fig:periodic-sources}
\end{figure*}

    Similar to CW, this category is a subset of static sources and can be removed with directional cuts. However, since it may be advantageous to keep those directions in some searches, an algorithm was developed to demonstrate removal based on timing alone.  
    We define a temporal test statistic ($\mathrm{TS}$) which gives a measure of the relative abundance of temporally nearby events with trigger times consistent with a period of $T$.  For each event, $i$, the difference in trigger times is calculated for a range of nearby events, indexed by $j$, within a specified time window $w$.  To allow for multiples of the period, we calculate the absolute difference to the nearest multiple of $T$ using:
    \begin{equation}
        d_{i,j} = \left| \left(t_i - t_j + \frac{T}{2}\right) \mathbin{\%} T - \frac{T}{2} \right|,
    \end{equation}
    where $\%$ refers to the floored modulo operation ($a\%n = a - n\lfloor\frac{a}{n}\rfloor$), resulting in $d_{i,j}$ being near zero for times close to an integer multiple of $T$.  Within each window $w$ containing $N_w(i)$ events, we construct a histogram with 20 bins ranging from 0 to $T/2$.  The top right portion of Figure~\ref{fig:periodic-sources} shows an example histogram, with red highlighting the bin, $c_{i,0}$, containing the $d_{i,j}$ most consistent with a periodicity of $T$ and green highlighting the 50\% of bins least consistent with $T$.  The test statistic for that event ($\mathrm{TS}_i$) is defined as the difference between the red region and the mean of the green region, given by:
    \begin{equation}
        \mathrm{TS}_{i} = c_{i,0} - \frac{1}{10} \sum_{k=10}^{19} c_{i,k}
    \end{equation}
    
    Arrival times from a uniform distribution would result in no significant difference in counts between the red and green regions, resulting in a $\mathrm{TS}$ near 0, while a set of perfectly periodic events would all lie within $c_{i,0}$, resulting a $\mathrm{TS}$ of 1.   Datasets contaminated with periodic noise sources are in-between these two extremes, resulting in a distribution of $\mathrm{TS}$ that is broadened when compared to uniform arrival times.  Periodic events can thus be highlighted from within a contaminated set of data by cutting on high $\mathrm{TS}$ values.  In Figure~\ref{fig:periodic-sources} we show how this algorithm can separate events observed arriving at a regular 60 Hz rate in data taken from September 2021.  This figure also shows how the baseline timing of the periodic events fluctuates with time as the 60 Hz drifts, which is handled by choosing a value for $w$ that is short relative to the fluctuations.  Depending on the desired efficiency, this algorithm will not flag all events arriving with a periodicity of $T$, but can isolate a clean subset of those events, which can then be used to motivate further targeted cuts based on template matching, direction, and signal properties to further improve the efficiency for removing this form of RFI.
    
\begin{figure*}[htbp]
\centering
\includegraphics[width=\iftoggle{double-column}{0\textwidth}{\textwidth}]{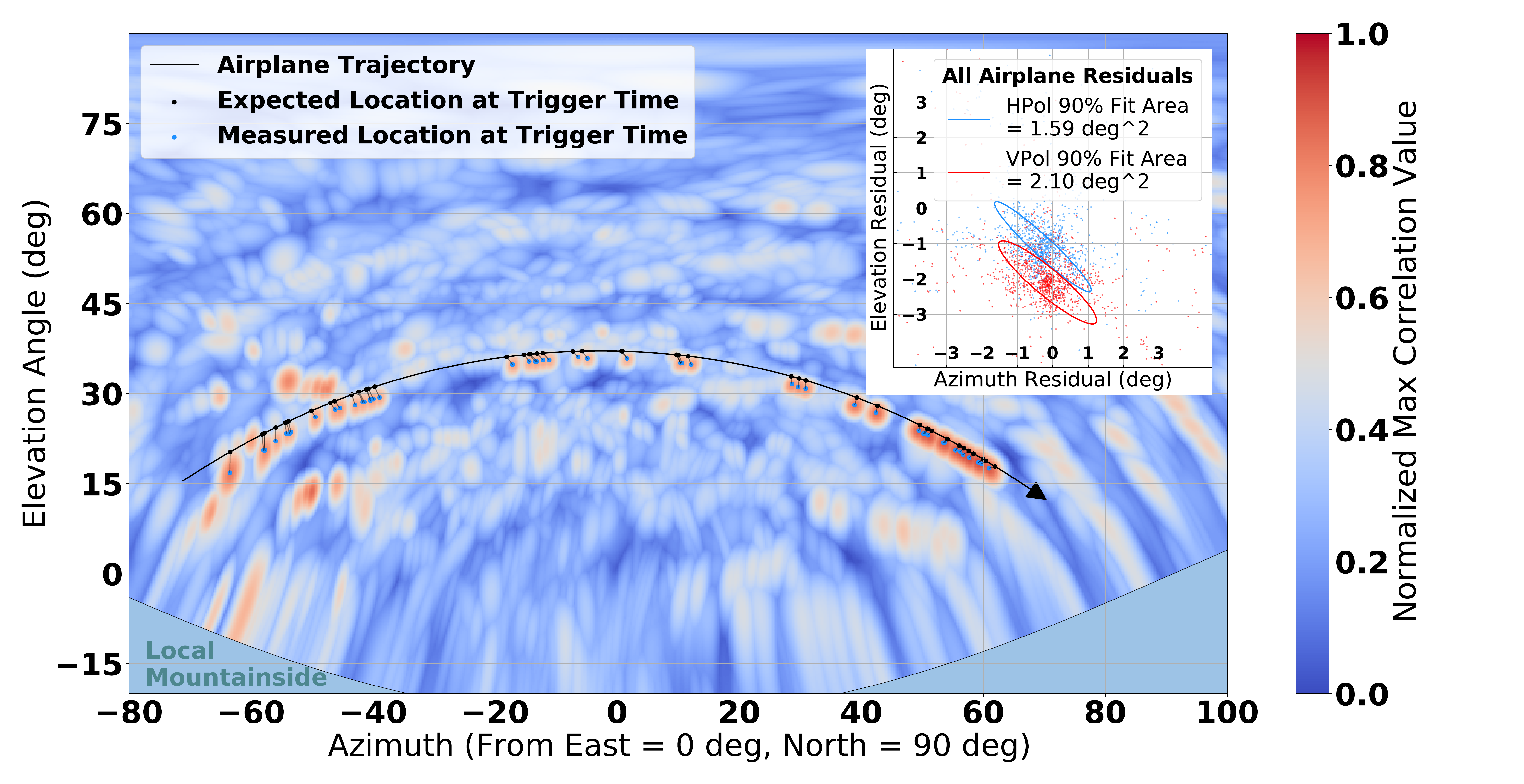}
\caption{Left: The stacked correlation map of 52~events corresponding to a single airplane track, with a colorscale corresponding to maximum correlation map value obtained from any event's individual map generated using all 8 channels.  The track of the corresponding airplane using ADS-B data obtained from The OpenSky Network~\cite{opensky} is shown with the black line, and spans $\sim$3.5 minutes.  The expected location of the airplane at the time of each triggered event in the map is shown with the black dots, and the measured location of the peak correlation value of each triggered event is shown with the blue dots.  Upper Right Inset:  Scatter plots showing the reconstruction offset observed for all airplanes when observed using either HPol (blue) or VPol (red) antennas, with a corresponding 2D Gaussian fit for each.  This plot demonstrates an observed systematic offset of approximately 1$^\circ$ in HPol and 2$^\circ$ in VPol (each polarization is calibrated independently).  This offset does not show significant angular or temporal dependence and is likely a result of the calibration.  Additionally, the 90\% integral area of the Gaussian fit observed for these events is larger than the \mbox{<\,0.1} sq.\@ degrees observed for static sources in Section~\ref{sec:calib}.  The original calibration was performed using mountainside pulsing and showed maximal reconstruction offsets of $\sim$1$^\circ$.  Future calibration campaigns using calibration sources mounted on drones would allow us to better constrain antenna positions by providing a large range of elevation angles for fitting and validation.  Calibration is discussed further in Section~\ref{sec:calib}.}
\label{fig:airplane}
\end{figure*}

  \textit{Airplanes}:  One of the few above horizon sources of RFI is airplanes.  As part of the above-horizon impulsive events search discussed in Section~\ref{sec:cr-discussion}, over 1000~RF triggered events were associated temporally and spatially with airplanes, corresponding to \mbox{>\,100} individual airplanes, an approximate observation rate of $\mathcal{O}(1)$ airplane per day.  Airplane signals have been identified by other experiments~\cite{PierreAuger:2015aqe}; though several potential sources have been described, no definitive cause for these signals has yet been determined.  The signals differ greatly in shape between airplanes and are not present for the majority of airplanes passing by the site.  Because of this we believe that the airplanes are not the source of these signals but rather serve as reflectors to signals originating on the ground. Therefore, we do not expect to see signals from all airplane tracks in our band. An example airplane-associated series of events is shown in Figure~\ref{fig:airplane}, which also shows the self-reported trajectory of the airplane superimposed~\cite{opensky}.  Apparent in this figure is a systematic offset in reconstruction direction for airplanes.  This offset is approximately 1$^\circ$ in HPol and 2$^\circ$ in VPol (where each polarization is calibrated independently).  This offset is small and does not significantly impact the results of this analysis, however understanding and fixing it is a priority for future analysis (see Section~\ref{sec:conclusion}).

\subsection{Above-Horizon Impulsive Events\label{sec:above-horizon-impulsive-search}}

We have categorized impulsive, above-horizon events in the prototype instrument data, identifying a variety of event classes~\cite{southallmoriond}.  We are especially interested in above-horizon impulsive events because they contain a sample of cosmic ray air shower events, which can be used to determine the nominal sensitivity of the prototype instrument.

We expect a few cosmic ray events per day to trigger the prototype given nominal beam voltage SNR thresholds of 5$\sigma$~\cite{Zeolla:2021bl}. Cosmic ray candidates will appear as isolated above-horizon events that are not identifiable as RFI events and do not cluster spatially or temporally. Additionally, cosmic ray events are expected to be impulsive signals that correlate well with cosmic ray templates from simulations. They will also have a polarization angle correlated with the source direction and the direction of radio emission from air showers in the local Earth's magnetic field (\textit{i.e.} $\sim\vec{v}\times \vec{B}$~\cite{Schroder:2016hrv}). 

Here we describe our classification process for above-horizon impulsive events, and show a likely cosmic ray candidate event from the data.  We present an analysis of $\sim$112 days of data taken from the beginning of September to the end of December 2021, consisting of $\sim$100 million RF-triggered events. 

\subsubsection{Identifying Above-Horizon Impulsive Events}
\label{sec:abovehorizon}
We first filter the data to remove both known frequencies of anthropogenic noise with static notch filters (at 27, 88.5, 107, 118, and 126 MHz in both polarizations, and additionally from 52.5 to 60.25 MHz in HPol channels, which removes RFI associated with the TV band).  We use a method called sine subtraction filtering~\cite{PhysRevD.93.122005}, where we filter the signals by fitting sine waves in the time domain with floating phase and amplitude, and remove any frequencies with amplitude above a threshold set in the analysis.  This method preserves causality in the data. We then remove the group delay added by the RF signal chain in the data to recover the original phase of the incident signals.

We then create a correlation map for each event and identify the most likely incident arrival direction for each by selecting the location of the peak cross-correlation value from one of three maps: 1) HPol channels only, 2) VPol channels only, and 3) the average of the two polarized maps.  We choose to use the peak location from the map that has the maximum  \textit{peak-to-sidelobe ratio} multiplied by the \textit{normalized map peak value}.  The peak-to-sidelobe ratio is the ratio of the main peak to the second brightest peak in the correlation map.  The \textit{normalized map peak value} is the ratio between the peak value and the optimal possible map value for that event, which would be obtained if a particular direction perfectly sampled the peak of each baseline's cross correlation.  Normalizing map peak values in this way counteracts the trend for low SNR events to have lower correlation values and thus lower map peak values.  For each map we mask out the direction of the mountainside itself (defined as the area below a simple plane fit to the antenna locations).

\begin{figure}[tbp]
\centering
\includegraphics[width=\iftoggle{double-column}{\columnwidth}{0.6\textwidth}]{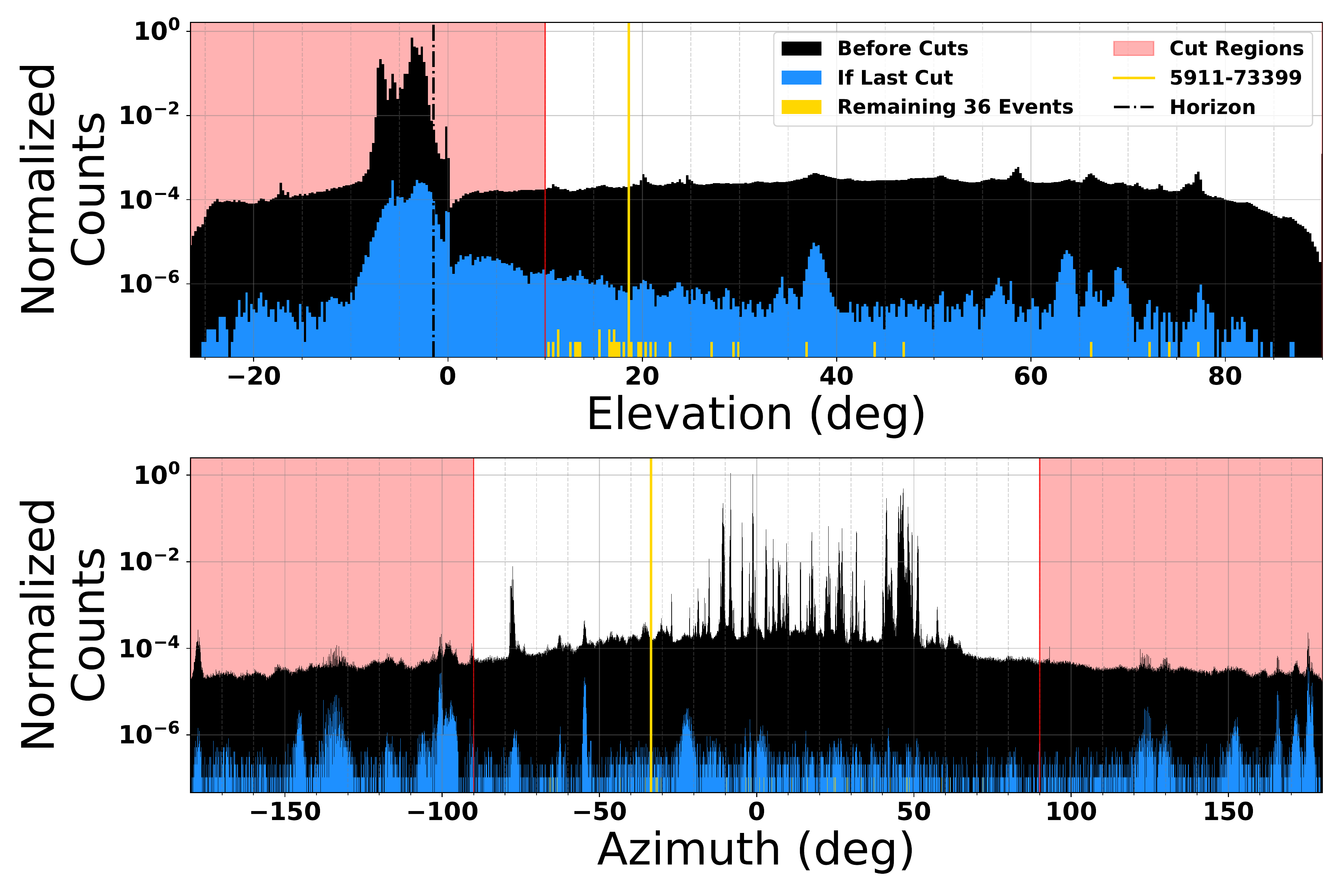}
\caption{Arrival directions of the received radio signal at the BEACON prototype for the full data set (black), the data set remaining after all other cuts have been applied (blue), and the 36 remaining events discussed in Section~\ref{sec:cr-discussion} (yellow). The reconstructed elevation (azimuth) for each event are shown in the top (bottom). Regions shown in red are excluded by the cut value placed at the red line.  For reference, the parameter values for the likely cosmic ray candidate event (discussed in Section~\ref{sec:cr-discussion}) is shown with the yellow vertical line (Event 5911-73399).  The approximate location of the horizon is shown on the top plot at an elevation angle of -1.5$^\circ$.}
\label{fig:directional-histograms}
\end{figure}

We then separate above-horizon from below-horizon events, which removes the vast majority of triggered events, which are dominated by static below-horizontal RFI sources (see Section~\ref{sec:rfi}). We keep events in our sample that have an arrival direction between $\left[-90^\circ, 90^\circ\right]$ in azimuth (East = $0^\circ$, North = $90^\circ$), and $\left[10^\circ, 90^\circ\right]$ in elevation as shown in Figure~\ref{fig:directional-histograms}.  The lower bound in elevation of $10^\circ$ above the horizontal is chosen to be far from the true horizon (which is $\sim$1.5$^\circ$ below the horizontal), to create a cleaner sample of downgoing events. The azimuthal cut restricts the search to the direction that the array is most sensitive to, which is to the East, since it is on an East-facing slope. Sources of RFI are finely resolved, suggesting that clustering could remove backgrounds in future searches.

We then develop a series of cut parameters to select for impulsive, isolated events that correlate with a cosmic ray template. We intentionally keep these cuts loose so any one cut is not overly restrictive, in order to investigate a variety of classes of events of interest above the horizon, while keeping a high fraction of triggered cosmic ray events in the remaining event sample. After all cuts, the data set is reduced to 5,440 events.  We list the cuts below, and in Table~\ref{tab:cuts}, along with the numbers and fraction of events that survive each cut.  The cuts are defined as:

\textit{Time Delay Clustering Cut}: Remove events that are in runs with more than 10 events that have the same measured arrival time delays between antenna pairs (with an absolute tolerance of 2.5 ns per baseline).  Runs in the data set are 1 to 3 hours long.  This cut is used to remove events that come from the same direction.
    
\textit{Peak-to-Sidelobe Ratio}:  Remove events for which the ratio of the main peak to the second brightest peak in the HPol and VPol correlation maps sums to less than 2.15.  A peak-to-sidelobe ratio near 1 indicates two peaks with comparable brightness.  This cut removes events where it is likely that the event could be mis-reconstructed, i.e. the main peak is indistinguishable from the sidelobes.
    
\textit{Impulsivity}: Remove events that have summed HPol and VPol impulsivities ($\mathcal{I}$) below 0.3.  $\mathcal{I}$ is a metric for measuring the impulsiveness of a signal~\cite{gorham2018constraints}, defined here as $\mathcal{I} = 2A-1$, where $A$ is the average of the cumulative distribution of fractional power contained within a 400 ns window centered on the peak of the Hilbert envelope of the aligned and averaged waveforms for a particular polarization. 
    
\begin{figure}[!t]
\centering
\includegraphics[width=\iftoggle{double-column}{\columnwidth}{0.65\textwidth}]{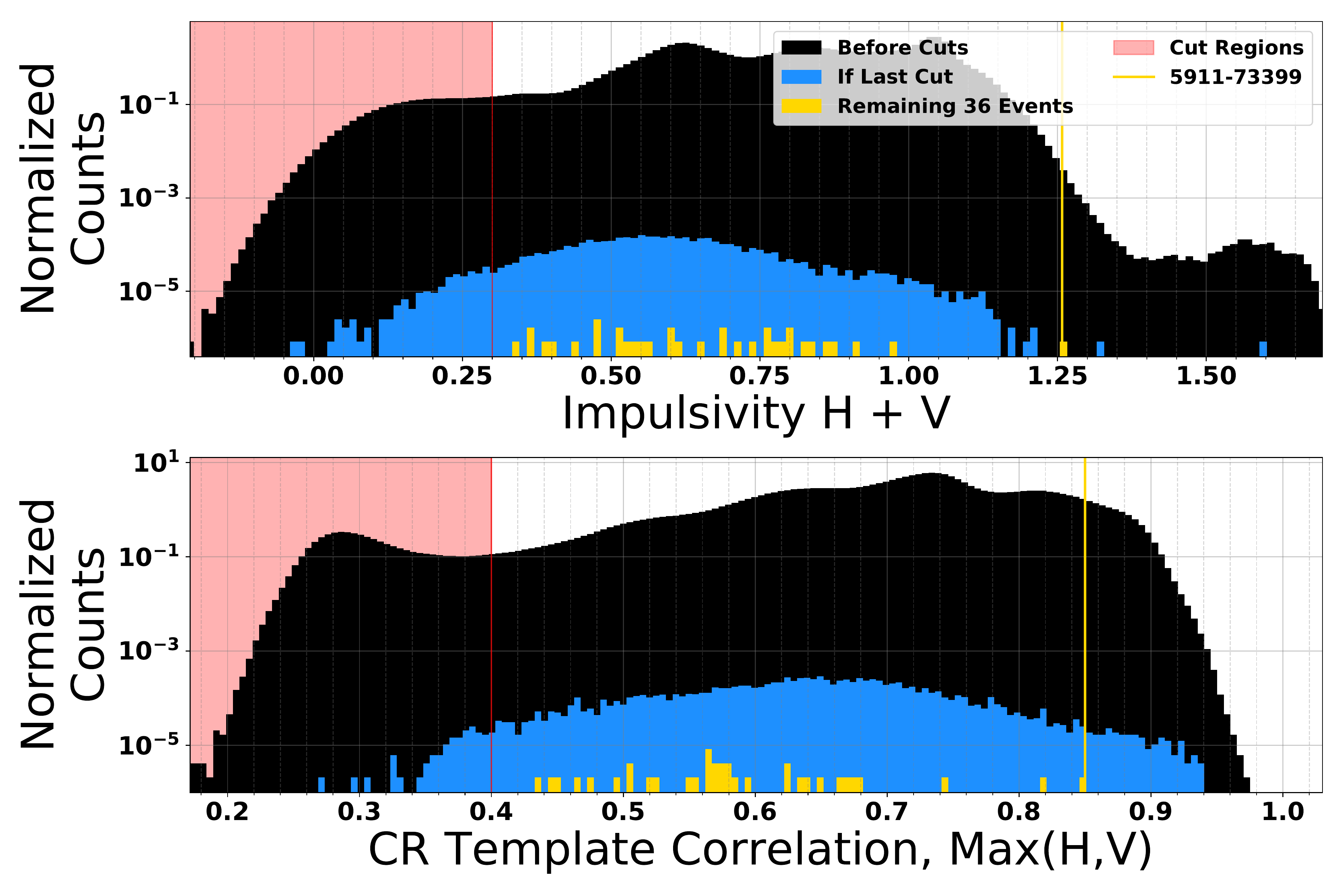}
\caption{Representative distributions of the impulsive character of the full data set (black), the data set remaining after all other cuts have been applied (blue), and the 36 remaining events discussed in Section~\ref{sec:cr-discussion} (yellow). The red line and region represents the cuts on both the combined impulsivity in  HPol and VPol channels and a correlation with a CR template. These cuts require the signal to be impulsive but are loose enough to allow for a variety of signal classes to classify above-horizon events.  For comparison, the parameter values for the likely cosmic ray candidate event (discussed in Section~\ref{sec:cr-discussion}) is shown with the yellow vertical line (Event 5911-73399).}
\label{fig:impulsive-histograms}
\end{figure}

\textit{Cosmic Ray Template Correlation}: Remove events for which neither polarization obtains a normalized correlation value of 0.4 with a simplified cosmic ray template.  The template used was a bipolar impulse with duration and amplitudes motivated by an off-axis angle of $1.37^\circ$ for a slightly upgoing air shower~\cite{zilles2020radio}.  This signal is then convolved with the appropriate channel-dependent responses of the prototype instrument, before undergoing the same filtration and cleaning as the waveforms, to create a template for correlation.
    
\textit{Likely Mis-Reconstructions of Known Below-Horizon Sources}: To remove events that mis-reconstruct above horizon due to prominent sidelobes of below horizon sources, a set of bright below horizon sources were identified.  Events are cut if their best below-horizon reconstruction direction is associated with a known RFI source. 
    
\textit{Signal Amplitude Differences}: Remove events that have significant peak-to-peak (P2P) voltage differences between HPol channels, where
the Max(P2P$_\mathrm{H}$) is 95 adu or more above the Min(P2P$_\mathrm{H}$).  This removes events where a subset of channels is significantly brighter than
the rest (indicative of local noise at the array or electronics issues), and small sample of events where one channel is not functioning properly.
    
\textit{Combined Normalized Map Peak Value}: Remove events that do not achieve a threshold percentage of their optimal achievable map value, using a combination of VPol and HPol maps.  We remove events where \mbox{$0.768  m_\mathrm{H} + 0.640  m_\mathrm{V} - 0.960$} is less than 0, where the normalized map peak value of the above-horizontal region in each polarization is $m$.
    
\textit{Combined Peak-To-Peak / (2 $\cdot$ Standard Deviation)}: Remove events where the signal amplitude (calculated as peak-to-peak divided by 2) is not sufficiently above the standard deviation of the observed ADC counts in that waveform.  Note that the standard deviation is calculated on the entire waveform, which includes the signal, so this metric is distinct from the SNR.  We remove events where the parameter \mbox{$0.878  r_\mathrm{H} + 0.479  r_\mathrm{V} - 5.267$} is less than 0, where $r$ is the ratio of half of the peak-to-peak over the standard deviation in each polarization.  

We show histograms of event distributions for a representative set of cut variables targeting impulsive events in Figure~\ref{fig:impulsive-histograms}, specifically for impulsivity and the correlation with a cosmic-ray template as defined above. The cuts in these two metrics were relatively loose, allowing us to investigate the varied signals we observe with the prototype. We highlight one event in particular that has a high value in both of these metrics. 

\begin{table*}[]
\begin{tabular}{lccc}
Cut Name   & Number of Events & Fraction Cut  & Fraction Cut  \\ 
& Remaining & Sequentially & if Applied First \\ \hline \hline
Full Data Set    & 96,483,288   &   &    \\ \hline
Elevation  & 1,830,144    & 0.98      & 0.98    \\
Azimuth         & 1,145,593     & 0.37        & 0.0075   \\
Time Delay Clustering, HPol   & 1,116,064    & 0.026    & 0.95  \\
Time Delay Clustering, VPol  & 1,104,002      & 0.011    & 0.85        \\
Peak-to-Sidelobe Ratio &  201,926          & 0.82   & 0.065        \\
Impulsivity  & 57,669  & 0.71   & 0.029     \\
Cosmic Ray Template Correlation & 42,184     & 0.27     & 0.028     \\
Associated with Below-Horizon Sources   & 38,274   & 0.093   & 0.79      \\
Signal Amplitude Differences   & 15,809     & 0.59         & 0.0038        \\
Combined Normalized Map Peak Value & 7,894  & 0.50    & 0.23                 \\
Combined Peak-to-Peak/(2 $\cdot$ Standard Deviation) & 5,440   & 0.31     & 0.044    \\ \hline \hline
\multicolumn{2}{l}{Hand-inspection breakdown of the 5,440 passing events:} & Number of Events & Fraction of Events \\ \hline
\multicolumn{2}{l}{Likely mis-reconstructions from below the horizon}  &  4,081  &  0.75 \\
\multicolumn{2}{l}{\hspace{0.7 cm} and Events with unstable amplifiers} & & \\
\multicolumn{2}{l}{Events associated with airplanes}  &  1,323  & 0.24  \\
\multicolumn{2}{l}{Remaining above-horizon events} &  36 &  0.0066  \\    
\end{tabular}
\caption{\label{tab:cuts}Summary of analysis cuts.  There are two stages of analysis: application of a variety of cuts (above the double line in the table) and a hand-inspection of events that pass those cuts (below the double line in the table).  The cut parameters and cut values used in the first stage of the analysis are described in Section~\ref{sec:abovehorizon}. The table shows the number of events remaining after each cut is applied sequentially, the fraction of events rejected by each cut when applied sequentially, and the fraction of events that are rejected if a given cut is applied first in the analysis.  The categorization of events by subsequent hand-inspection of the passing 5,440 events is also shown.} 
\end{table*} 

\subsubsection{\label{sec:cr-discussion} Remaining Above-Horizon Events}

The remaining 5,440 events were inspected by hand.  We found that three broad categories of events remained, as shown in Table~\ref{tab:cuts}.  This hand-categorization of impulsive events that appear to come from above the horizon is important for understanding the RFI environment of the BEACON prototype site, to inform future design decisions
and future analyses of the data.  Events were categorized into three broad categories: likely mis-reconstructions of RFI that originates from below the horizon, events associated with airplanes, and other impulsive above-horizon events.

\textit{Likely Mis-reconstructions of below-horizon events and events with amplifier instability:} The largest category is events that are likely to be mis-reconstructions of below-horizon sources of RFI and events that exhibit instability in the amplifier chain, constituting 75\% of the data set that passed all cuts described in Section~\ref{sec:abovehorizon}.  The vast majority of triggered events originate from below the horizon, as shown in Table~\ref{tab:cuts}, and if the correlation map peaks on a true sidelobe of the signal, it is possible for such events to appear to come from above the horizontal and pass the elevation cut applied to the data.  Manual inspection of the correlation map can identify these events.  Additionally, events are identified that have features in the data that are a result of instability in the amplifiers used in the electronics chain as well as events containing multiple impulses which can create false cross-correlations above the horizon.

\textit{Events associated with airplane tracks: }The next largest category is events that were associated temporally and spatially with over~100 known airplane trajectories from The OpenSky Network~\cite{opensky}, which contains an extensive database of ADS-B airplane data that most airplanes are required to transmit~\cite{CFR-ADS-B-91.225,CFR-ADS-B-91.227}. An example airplane track seen in the data is shown in Figure~\ref{fig:airplane}.  64 individual airplanes were associated with at least four triggered events, and six airplanes caused 50 or more triggered events.  Additional events created other temporally clustered trajectories across the sky, but with no known corresponding airplane track in the database; these events have also been tagged as likely airplane events.  This category of events constitutes 24\% of the sample.

\begin{figure*}[p!]
    \centering
    \includegraphics[width=\iftoggle{double-column}{0.95\textwidth}{0.85\textwidth}]{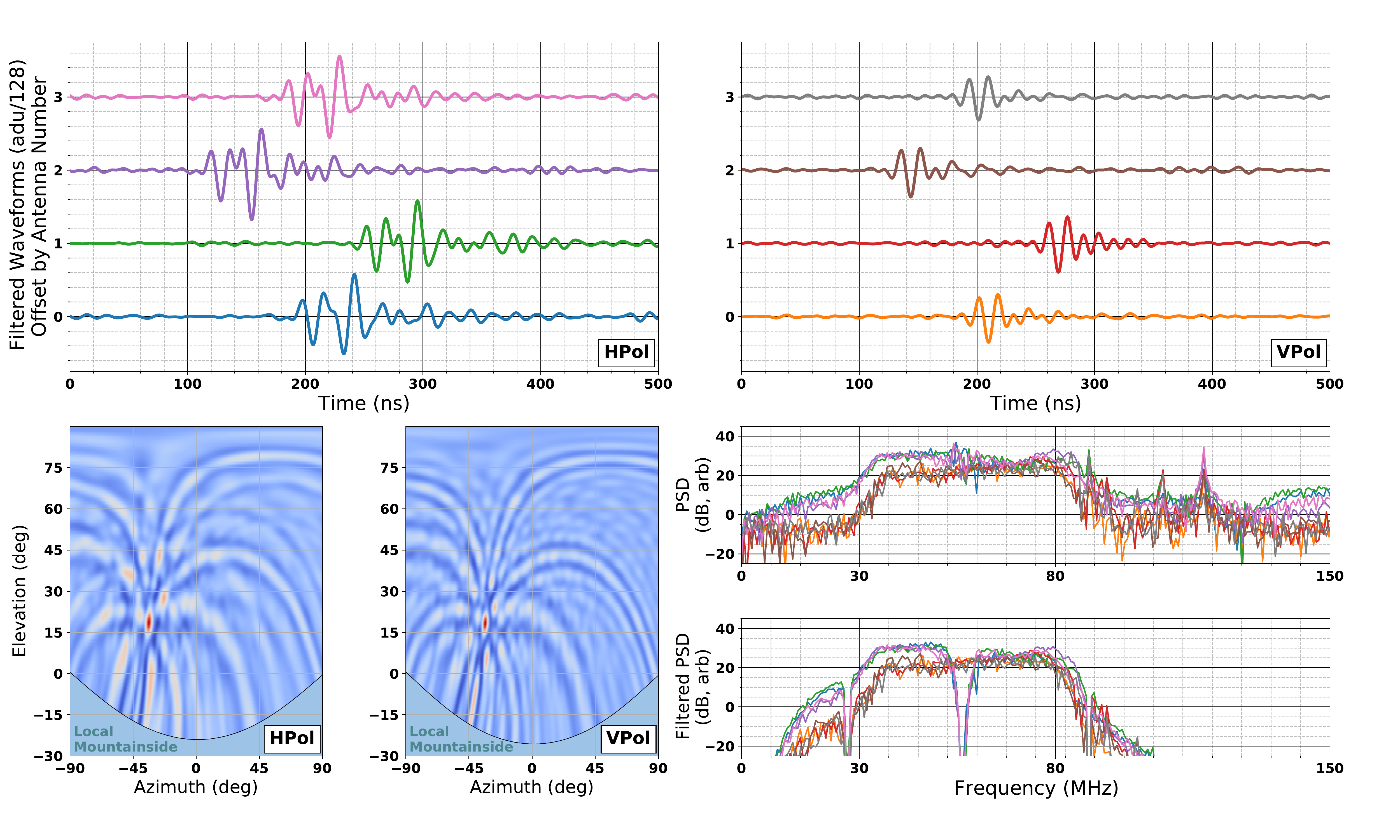}
    \caption{Event display for a likely cosmic ray event (Event 5911-73399).  Top: Waveforms from each of the 8~channels, normalized and offset such that the y-scale indicates the antenna number for each waveform.  This event has an averaged single-channel voltage SNR of 42.5 in HPol and 38.6 in VPol.   Bottom Left: HPol and VPol correlation maps. The colorscale of each map is individually normalized, and the region of the maps pointing into the local mountainside is masked out.  Bottom Right: The Power Spectral Density (PSD) before and after filtering.  The data has been filtered as described in Section~\ref{sec:abovehorizon}.}
    \label{fig:candidate_full}
\end{figure*}

\textit{Remaining events:} Of the 5,440 events which passed the cuts aimed to identify impulsive above-horizon signals, only 36 (less than 1\%) were not associated with airplane tracks and were not categorized as likely mis-reconstructions of below-horizon events or events with unstable electronics; parameter distributions of these events are included in Figures~\ref{fig:directional-histograms} and \ref{fig:impulsive-histograms}. The events are uniform in azimuth and show some structure in the elevation angle. The structure could be consistent with either sidelobes from below-horizon sources or cosmic rays, which are expected to be highly inclined on average for the BEACON geometry. Understanding this distribution will be the subject of future analyses (Section~\ref{sec:future}). One event of interest from this sample is shown in Figure~\ref{fig:candidate_full} (event 5911-73399), and is a likely cosmic ray event.  The remaining events are of as yet unknown origin and will be the subject of future study; these events are impulsive and above horizon and may include a combination of unidentified backgrounds and additional cosmic ray events.

The candidate cosmic ray event has the third highest impulsivity of all 5,440 events that pass cuts, and the highest among the 36 remaining events. Further inspection of the two events with higher impulsivity categorized them as a likely mis-reconstruction of a below-horizon event and a likely airplane event. The candidate event also has the highest SNR (beam voltage SNR of 91\,$\sigma$ in HPol; 58\,$\sigma$ in VPol for the processed waveforms), peak-to-sidelobe ratio (\mbox{$>$\,1.7} for each polarization), and template correlation values (\mbox{$>$\,0.83} for each polarization) among the 36 remaining events.  This event also does not occur during a time of significant lightning activity.

\begin{figure*}[p!]
    \centering
    \includegraphics[width=\iftoggle{double-column}{0.95\textwidth}{0.85\textwidth}]{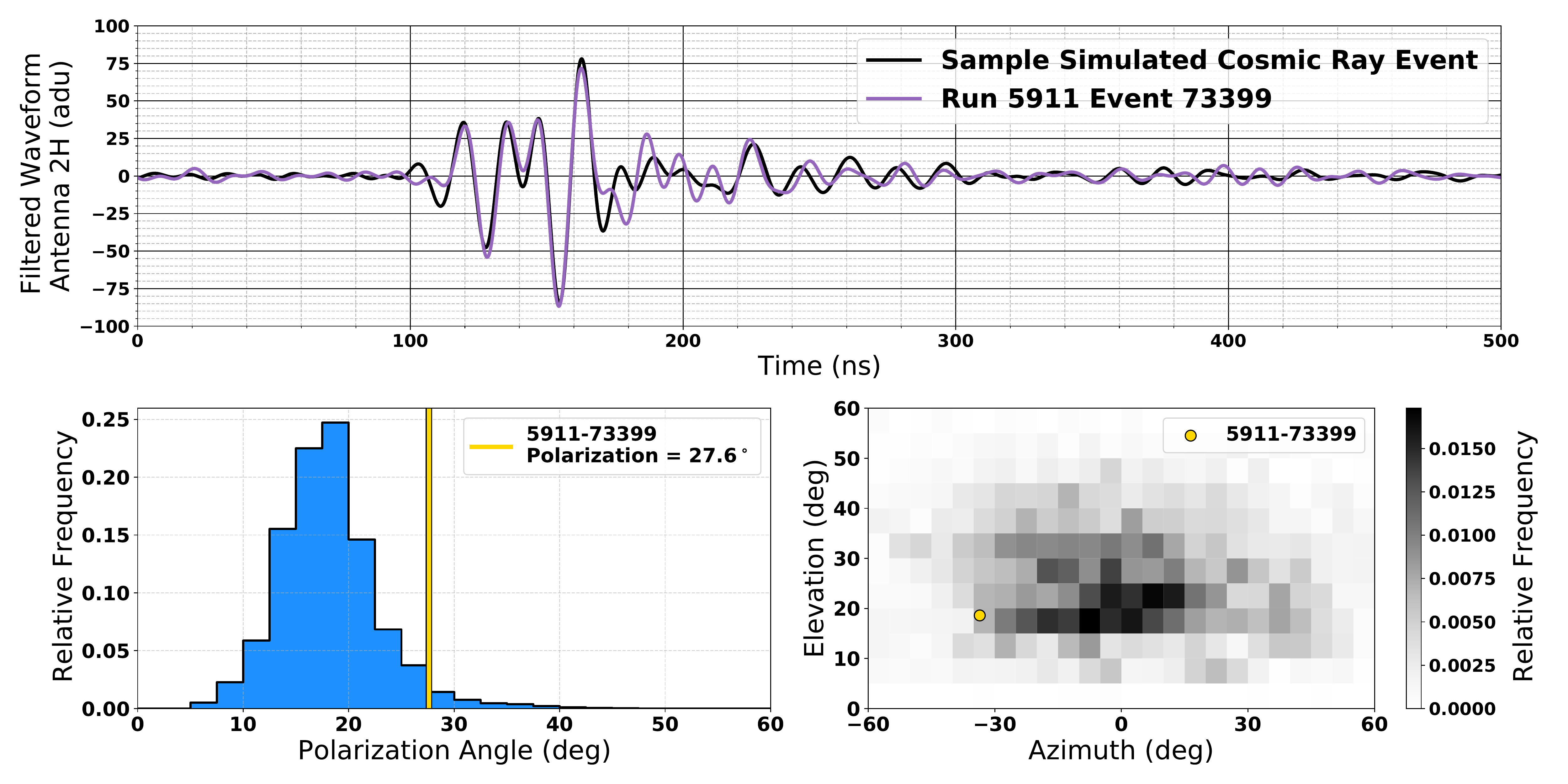}
    \caption{Top: The waveform for Event 5911-73399 from Antenna 2H superimposed with a sample simulated cosmic ray signal with realistic thermal noise levels~\cite{Zeolla:2021bl}, which has been convolved with the system response of the same channel.  Both waveforms have been filtered as described in Section~\ref{sec:abovehorizon}.  Bottom Left: The distribution of expected observed linear polarization angles for triggered simulated events.  The polarization angle of the cosmic ray candidate event is shown with a yellow line. The measured polarization angle of $\sim$28$^{\circ}$ is consistent with the purely geometric expectation of $\sim$30$^{\circ}$, calculated assuming a geomagnetic signal arriving from the appropriate arrival direction and local magnetic field.  Bottom Right: The distribution of expected azimuth and elevation for simulated events compared to the candidate cosmic ray event (in yellow).}
    \label{fig:candidate_template}
\end{figure*}

Figure~\ref{fig:candidate_template} shows the waveform from the event of interest alongside an event waveform generated with the cosmic ray simulation~\cite{Zeolla:2021bl} and compares the observed linear polarization angle and arrival direction with simulated distributions.  The tangent of the polarization angle is calculated as the ratio of the maximum of the aligned and averaged de-dispersed waveforms in VPol to HPol when upsampled and symmetric filtering is applied across polarizations (such that VPol is filtered with the TV notch filter as well, ensuring similar power is lost in both averaged waveforms and a representative ratio is preserved).  In this way the measured polarization angle of $\sim$28$^{\circ}$ is consistent with the purely geometric expectation of $\sim$30$^{\circ}$, with an uncertainty in the polarization measurement of $\sim$2$^{\circ}$, corresponding to the $\sim$10\% observed variance in gain matching among channels.  The geomagnetic expectation is for a signal arriving from the appropriate arrival direction and local magnetic field for this event. 

\subsubsection{Future Work}\label{sec:future}
The categorization of impulsive above-horizon events in the prototype instrument data set is a critical step in defining cuts for future cosmic ray searches.  While informative for this analysis, the hand-inspection of events after cuts are applied indicates that additional automated cuts would need to be made to perform a true cosmic ray search. For example, the structure in the spatial distribution of events seen above the horizon (e.g.\ the elevation distribution in Figure~\ref{fig:directional-histograms}) indicates that a set of clustering cuts to remove events associated with below-horizon sources would be effective.  We are planning further analyses that will leverage our understanding of the prototype system and local RFI sources to perform a cosmic ray search with the data. These searches will benefit from search metrics that are efficient at removing backgrounds and identifying cosmic ray events with low SNR. Cuts that take advantage of the directional and temporal clustering in RFI sources may be sensitive to weaker signals; however, confidence in identification can be improved when clustering is combined with cuts that emphasize the impulsive characteristics of cosmic-ray signals. Additionally, as discussed in Section \ref{sec:cr-discussion}, the polarization angle of air shower events is a predictable metric dependent on local magnetic field and signal arrival direction.  In the impulsive above horizon search presented here we use this only as a check on our final candidate, however in future searches polarization angle will serve as a powerful metric for automated cuts. Combining the results of that search with input from the cosmic ray simulation will lead to an updated sensitivity estimate to tau neutrinos of the full-scale BEACON array.

\section{\label{sec:conclusion}Conclusions}

The BEACON prototype instrument has been in operation since 2018.  The current station design is robust and with its custom antennas and phased array trigger represents important first steps towards a scalable implementation of the full BEACON array.

We have used data from the prototype instrument to verify the performance of the array and understand the RFI environment at the BEACON prototype site.  We have developed analysis techniques to identify above horizon RFI sources such as airplanes, and to isolate events consistent with the expected properties of a cosmic ray.  The results of this analysis have already validated the phased trigger's ability to maintain sensitivity to above horizon events using a small number of antennas in a noise-dominated environment like the Californian and Nevada deserts. While the RFI rates at the current prototype site are higher than would be beneficial for a larger instrument, the environment provides an important stress test of the trigger's capabilities. 

The next stage for the BEACON prototype is to develop a full cosmic ray search trained on simulated data and building on the background studies presented here and on techniques from other autonomous searches for radio signals from air showers. The dominant source of backgrounds come from below the horizon and are well clustered both spatially and temporally, suggesting that they may be removable as has been done in prior searches~\cite{monroe2020self, charrier2019autonomous, Schoorlemmer:2015afa}. We plan to conduct a template search based on simulated radio emission from cosmic rays and exploiting clustering cuts. 

Since anthropogenic noise predominantly comes from below the horizon and constitutes the main source of background, there will be a need for more background rejection power in a search for upgoing tau neutrinos compared to downgoing cosmic rays.  This ultimately could translate to a loss in analysis efficiency for a given background rate. However, we may be able to further exploit differences in the characteristics of the signals -- their spectra, their isotropy, and impulse response -- relative to the backgrounds. We can also further tune the beamforming trigger to down-weight or directionally mask regular sources of RFI at a given site. These studies will be important to pursue in future works.

The concept for each BEACON station includes more antennas and longer baselines than the prototype and could therefore achieve lower thresholds and improved background rejection.  Measurements of correlation map characteristics, SNR, pointing resolution, and trigger thresholds all benefit from the additional antennas and longer baselines of a full station, enabling better separation of below-horizon and above-horizon events.  Additionally, multiple stations with differing views of overlapping effective volumes can be used to veto anthropogenic noise, compared to air shower signals, which are highly beamed. 

Finally, we note that we are exploring hardware upgrades for the prototype. Antenna position calibration and trigger validation using an RF source mounted on a drone can enable a more complete calibration of the in-situ antenna beam patterns~\cite{Nam:2021gcb}.  This drone pulser will also be used to further understand the observed elevation offset in above-horizon events from airplanes (see Figure~\ref{fig:airplane}), and determine whether this offset is intrinsic to the hardware, current calibration, or in our interpretation of the airplane database, which is important for trusting above-horizon reconstruction accuracy in future analysis.  While not cost-effective for a full-scale detector, adding scintillators to the prototype can improve cosmic ray identification and validation at the prototype stage.  Signals from scintillator detectors could be digitized alongside the existing RF channels and serve to validate RF-only triggered events~\cite{Mulrey:2021fqw}.  Finally, an updated DAQ is being designed, which because it is modular and flexible, can form the basis of autonomous stations with more antennas. This will allow us to scale the BEACON detector to the hundreds or thousands of stations needed to detect the tau neutrino flux.

%\appendix

%\section*{References}
\clearpage
\pagebreak
%\newpage
%\clearpage
\bibliography{bib}% Produces the bibliography via BibTeX.

\providecommand{\noopsort}[1]{}\providecommand{\singleletter}[1]{#1}%
\begin{thebibliography}{10}
\expandafter\ifx\csname url\endcsname\relax
  \def\url#1{\texttt{#1}}\fi
\expandafter\ifx\csname urlprefix\endcsname\relax\def\urlprefix{URL }\fi
\expandafter\ifx\csname href\endcsname\relax
  \def\href#1#2{#2} \def\path#1{#1}\fi

\bibitem{Aab_2020}
A.~Aab, et~al., {Measurement of the cosmic-ray energy spectrum above
  $2.5\ifmmode\times\else\texttimes\fi{}{10}^{18}\text{ }\text{ }\mathrm{eV}$
  using the Pierre Auger Observatory}, Phys. Rev. D 102 (2020) 062005.
\newblock \href {http://dx.doi.org/10.1103/PhysRevD.102.062005}
  {\path{doi:10.1103/PhysRevD.102.062005}}.

\bibitem{Abu_Zayyad_2013}
T.~Abu-Zayyad, et~al., {The Cosmic-Ray Energy Spectrum Observed with the
  Surface Detector of the Telescope Array Experiment}, The Astrophysical
  Journal 768~(1) (2013) L1.
\newblock \href {http://dx.doi.org/10.1088/2041-8205/768/1/l1}
  {\path{doi:10.1088/2041-8205/768/1/l1}}.

\bibitem{ABBASI200953}
R.~Abbasi, et~al., {Measurement of the flux of ultra high energy cosmic rays by
  the stereo technique}, Astroparticle Physics 32~(1) (2009) 53--60.
\newblock \href {http://dx.doi.org/10.1016/j.astropartphys.2009.06.001}
  {\path{doi:10.1016/j.astropartphys.2009.06.001}}.

\bibitem{greisen1966end}
K.~Greisen, {End to the Cosmic Ray Spectrum?}, Phys. Rev. Lett 16 (1966)
  748--750.

\bibitem{zatsepin1966upper}
G.~Zatsepin, V.~Kuz'min, {Upper limit of the spectrum of cosmic rays}, Soviet
  Journal of Experimental and Theoretical Physics Letters 4 (1966) 78.

\bibitem{Berezinsky:1969erk}
V.~S. Berezinsky, G.~T. Zatsepin, {Cosmic rays at ultrahigh-energies
  (neutrino?)}, Phys. Lett. B 28 (1969) 423--424.
\newblock \href {http://dx.doi.org/10.1016/0370-2693(69)90341-4}
  {\path{doi:10.1016/0370-2693(69)90341-4}}.

\bibitem{PierreAuger:2016use}
A.~Aab, et~al., {Combined fit of spectrum and composition data as measured by
  the Pierre Auger Observatory}, JCAP 04 (2017) 038, [Erratum: JCAP 03, E02
  (2018)].
\newblock \href {http://arxiv.org/abs/1612.07155} {\path{arXiv:1612.07155}},
  \href {http://dx.doi.org/10.1088/1475-7516/2017/04/038}
  {\path{doi:10.1088/1475-7516/2017/04/038}}.

\bibitem{Fang:2013cba}
K.~Fang, K.~Kotera, A.~V. Olinto, {Ultrahigh Energy Cosmic Ray Nuclei from
  Extragalactic Pulsars and the effect of their Galactic counterparts}, JCAP 03
  (2013) 010.
\newblock \href {http://arxiv.org/abs/1302.4482} {\path{arXiv:1302.4482}},
  \href {http://dx.doi.org/10.1088/1475-7516/2013/03/010}
  {\path{doi:10.1088/1475-7516/2013/03/010}}.

\bibitem{AlvesBatista:2019tlv}
R.~Alves~Batista, et~al., {Open Questions in Cosmic-Ray Research at Ultrahigh
  Energies}, Front. Astron. Space Sci. 6 (2019) 23.
\newblock \href {http://arxiv.org/abs/1903.06714} {\path{arXiv:1903.06714}},
  \href {http://dx.doi.org/10.3389/fspas.2019.00023}
  {\path{doi:10.3389/fspas.2019.00023}}.

\bibitem{Coleman:2022abf}
A.~Coleman, et~al., {Ultra-High-Energy Cosmic Rays: The Intersection of the
  Cosmic and Energy Frontiers}\href {http://arxiv.org/abs/2205.05845}
  {\path{arXiv:2205.05845}}.

\bibitem{IceCube:2013low}
M.~G. Aartsen, et~al., {Evidence for High-Energy Extraterrestrial Neutrinos at
  the IceCube Detector}, Science 342 (2013) 1242856.
\newblock \href {http://arxiv.org/abs/1311.5238} {\path{arXiv:1311.5238}},
  \href {http://dx.doi.org/10.1126/science.1242856}
  {\path{doi:10.1126/science.1242856}}.

\bibitem{IceCube:2020acn}
M.~G. Aartsen, et~al., {Characteristics of the Diffuse Astrophysical Electron
  and Tau Neutrino Flux with Six Years of IceCube High Energy Cascade Data},
  Phys. Rev. Lett. 125~(12) (2020) 121104.
\newblock \href {http://arxiv.org/abs/2001.09520} {\path{arXiv:2001.09520}},
  \href {http://dx.doi.org/10.1103/PhysRevLett.125.121104}
  {\path{doi:10.1103/PhysRevLett.125.121104}}.

\bibitem{IceCube:2020wum}
R.~Abbasi, et~al., {The IceCube high-energy starting event sample: Description
  and flux characterization with 7.5 years of data}, Phys. Rev. D 104 (2021)
  022002.
\newblock \href {http://arxiv.org/abs/2011.03545} {\path{arXiv:2011.03545}},
  \href {http://dx.doi.org/10.1103/PhysRevD.104.022002}
  {\path{doi:10.1103/PhysRevD.104.022002}}.

\bibitem{doi:10.1126/science.aat1378}
{MAGIC, AGILE, ASAS-SN, HAWC, HESS, INTEGRAL, Kanata, Kiso, Kapteyn, Liverpool
  Telescope, Subaru, Swift/NuSTAR, VERITAS, and VLA/17B-403 teams, and others},
  {Multimessenger observations of a flaring blazar coincident with high-energy
  neutrino IceCube-170922A}, Science 361~(6398) (2018) eaat1378.
\newblock \href {http://dx.doi.org/10.1126/science.aat1378}
  {\path{doi:10.1126/science.aat1378}}.

\bibitem{Choubey:2009jq}
S.~Choubey, W.~Rodejohann, {Flavor Composition of UHE Neutrinos at Source and
  at Neutrino Telescopes}, Phys. Rev. D 80 (2009) 113006.
\newblock \href {http://arxiv.org/abs/0909.1219} {\path{arXiv:0909.1219}},
  \href {http://dx.doi.org/10.1103/PhysRevD.80.113006}
  {\path{doi:10.1103/PhysRevD.80.113006}}.

\bibitem{Bustamante:2015waa}
M.~Bustamante, J.~F. Beacom, W.~Winter, {Theoretically Palatable Flavor
  Combinations of Astrophysical Neutrinos}, Phys. Rev. Lett. 115~(16) (2015)
  161302.
\newblock \href {http://arxiv.org/abs/1506.02645} {\path{arXiv:1506.02645}},
  \href {http://dx.doi.org/10.1103/PhysRevLett.115.161302}
  {\path{doi:10.1103/PhysRevLett.115.161302}}.

\bibitem{Pakvasa:2007dc}
S.~Pakvasa, W.~Rodejohann, T.~J. Weiler, {Flavor Ratios of Astrophysical
  Neutrinos: Implications for Precision Measurements}, JHEP 02 (2008) 005.
\newblock \href {http://arxiv.org/abs/0711.4517} {\path{arXiv:0711.4517}},
  \href {http://dx.doi.org/10.1088/1126-6708/2008/02/005}
  {\path{doi:10.1088/1126-6708/2008/02/005}}.

\bibitem{Song:2020nfh}
N.~Song, S.~W. Li, C.~A. Arg\"uelles, M.~Bustamante, A.~C. Vincent, {The Future
  of High-Energy Astrophysical Neutrino Flavor Measurements}, JCAP 04 (2021)
  054.
\newblock \href {http://arxiv.org/abs/2012.12893} {\path{arXiv:2012.12893}},
  \href {http://dx.doi.org/10.1088/1475-7516/2021/04/054}
  {\path{doi:10.1088/1475-7516/2021/04/054}}.

\bibitem{Bustamante:2019sdb}
M.~Bustamante, M.~Ahlers, {Inferring the Flavor of High-Energy Astrophysical
  Neutrinos at Their Sources}, Phys. Rev. Lett. 122~(24) (2019) 241101.
\newblock \href {http://arxiv.org/abs/1901.10087} {\path{arXiv:1901.10087}},
  \href {http://dx.doi.org/10.1103/PhysRevLett.122.241101}
  {\path{doi:10.1103/PhysRevLett.122.241101}}.

\bibitem{Xing:2006uk}
Z.-Z. Xing, S.~Zhou, {Towards Determination of the Initial Flavor Composition
  of Ultrahigh-energy Neutrino Fluxes with Neutrino Telescopes}, Phys. Rev. D
  74 (2006) 013010.
\newblock \href {http://arxiv.org/abs/astro-ph/0603781}
  {\path{arXiv:astro-ph/0603781}}, \href
  {http://dx.doi.org/10.1103/PhysRevD.74.013010}
  {\path{doi:10.1103/PhysRevD.74.013010}}.

\bibitem{Valera:2022ylt}
V.~B. Valera, M.~Bustamante, C.~Glaser, {The ultra-high-energy neutrino-nucleon
  cross section: measurement forecasts for an era of cosmic EeV-neutrino
  discovery }\href {http://arxiv.org/abs/2204.04237} {\path{arXiv:2204.04237}}.

\bibitem{Denton:2020jft}
P.~B. Denton, Y.~Kini, {Ultra-High-Energy Tau Neutrino Cross Sections with
  GRAND and POEMMA}, Phys. Rev. D 102 (2020) 123019.
\newblock \href {http://arxiv.org/abs/2007.10334} {\path{arXiv:2007.10334}},
  \href {http://dx.doi.org/10.1103/PhysRevD.102.123019}
  {\path{doi:10.1103/PhysRevD.102.123019}}.

\bibitem{Connolly:2011vc}
A.~Connolly, R.~S. Thorne, D.~Waters, {Calculation of High Energy
  Neutrino-Nucleon Cross Sections and Uncertainties Using the MSTW Parton
  Distribution Functions and Implications for Future Experiments}, Phys. Rev. D
  83 (2011) 113009.
\newblock \href {http://arxiv.org/abs/1102.0691} {\path{arXiv:1102.0691}},
  \href {http://dx.doi.org/10.1103/PhysRevD.83.113009}
  {\path{doi:10.1103/PhysRevD.83.113009}}.

\bibitem{Esteban:2022uuw}
I.~Esteban, S.~Prohira, J.~F. Beacom, {Detector Requirements for
  Model-Independent Measurements of Ultrahigh Energy Neutrino Cross Sections
  }\href {http://arxiv.org/abs/2205.09763} {\path{arXiv:2205.09763}}.

\bibitem{wissel2020prospects}
S.~Wissel, A.~Romero-Wolf, H.~Schoorlemmer, W.~R.~C. Jr.,
  J.~Alvarez-Mu{\~{n}}iz, E.~Zas, A.~Cummings, C.~Deaconu, K.~Hughes,
  A.~Ludwig, J.~Morancy, E.~Oberla, C.~Paciaroni, S.~Prohira, D.~Southall,
  M.~Stapel-Kalat, B.~Strutt, M.~Vasquez, A.~Vieregg, {Prospects for
  High-Elevation Radio Detection of > 100 {PeV} Tau Neutrinos}, Journal of
  Cosmology and Astroparticle Physics 2020~(11) (2020) 065--065.
\newblock \href {http://dx.doi.org/10.1088/1475-7516/2020/11/065}
  {\path{doi:10.1088/1475-7516/2020/11/065}}.

\bibitem{Fargion:1999se}
D.~Fargion, A.~Aiello, R.~Conversano, {Horizontal tau air showers from
  mountains in deep valley: Traces of UHECR neutrino tau}, in: Proceedings,
  26th International Cosmic Ray Conference, August 17-25, 1999, Salt Lake City:
  Invited, Rapporteur, and Highlight Papers, 1999, p. 396, [2,396(1999)].
\newblock \href {http://arxiv.org/abs/astro-ph/9906450}
  {\path{arXiv:astro-ph/9906450}}.

\bibitem{Feng:2001ue}
J.~L. Feng, P.~Fisher, F.~Wilczek, T.~M. Yu, {Observability of earth skimming
  ultrahigh-energy neutrinos}, Phys. Rev. Lett. 88 (2002) 161102.
\newblock \href {http://arxiv.org/abs/hep-ph/0105067}
  {\path{arXiv:hep-ph/0105067}}, \href
  {http://dx.doi.org/10.1103/PhysRevLett.88.161102}
  {\path{doi:10.1103/PhysRevLett.88.161102}}.

\bibitem{Zas:2005zz}
E.~Zas, {Neutrino detection with inclined air showers}, New J. Phys. 7 (2005)
  130.
\newblock \href {http://arxiv.org/abs/astro-ph/0504610}
  {\path{arXiv:astro-ph/0504610}}, \href
  {http://dx.doi.org/10.1088/1367-2630/7/1/130}
  {\path{doi:10.1088/1367-2630/7/1/130}}.

\bibitem{alvarez2012monte}
J.~Alvarez-Mu{\~n}iz, W.~Carvalho~Jr, E.~Zas, {Monte Carlo simulations of radio
  pulses in atmospheric showers using ZHAireS}, Astroparticle Physics 35~(6)
  (2012) 325--341.
\newblock \href {http://dx.doi.org/10.1016/j.astropartphys.2011.10.005}
  {\path{doi:10.1016/j.astropartphys.2011.10.005}}.

\bibitem{Schoorlemmer:2015afa}
H.~Schoorlemmer, et~al., {Energy and Flux Measurements of Ultra-High Energy
  Cosmic Rays Observed During the First ANITA Flight}, Astropart. Phys. 77
  (2016) 32--43.
\newblock \href {http://arxiv.org/abs/1506.05396} {\path{arXiv:1506.05396}},
  \href {http://dx.doi.org/10.1016/j.astropartphys.2016.01.001}
  {\path{doi:10.1016/j.astropartphys.2016.01.001}}.

\bibitem{LOPES:2005ipv}
H.~Falcke, et~al., {Detection and imaging of atmospheric radio flashes from
  cosmic ray air showers}, Nature 435 (2005) 313--316.
\newblock \href {http://arxiv.org/abs/astro-ph/0505383}
  {\path{arXiv:astro-ph/0505383}}, \href
  {http://dx.doi.org/10.1038/nature03614} {\path{doi:10.1038/nature03614}}.

\bibitem{Ardouin:2009zp}
D.~Ardouin, et~al., {Geomagnetic origin of the radio emission from cosmic ray
  induced air showers observed by CODALEMA}, Astropart. Phys. 31 (2009)
  192--200.
\newblock \href {http://arxiv.org/abs/0901.4502} {\path{arXiv:0901.4502}},
  \href {http://dx.doi.org/10.1016/j.astropartphys.2009.01.001}
  {\path{doi:10.1016/j.astropartphys.2009.01.001}}.

\bibitem{Schellart:2013bba}
P.~Schellart, et~al., {Detecting cosmic rays with the LOFAR radio telescope},
  Astron. Astrophys. 560 (2013) A98.
\newblock \href {http://arxiv.org/abs/1311.1399} {\path{arXiv:1311.1399}},
  \href {http://dx.doi.org/10.1051/0004-6361/201322683}
  {\path{doi:10.1051/0004-6361/201322683}}.

\bibitem{monroe2020self}
R.~Monroe, et~al., {Self-triggered radio detection and identification of cosmic
  air showers with the OVRO-LWA}, Nuclear Instruments and Methods in Physics
  Research Section A: Accelerators, Spectrometers, Detectors and Associated
  Equipment 953 (2020) 163086.
\newblock \href {http://dx.doi.org/10.1016/j.nima.2019.163086}
  {\path{doi:10.1016/j.nima.2019.163086}}.

\bibitem{Bezyazeekov:2018yjw}
P.~A. Bezyazeekov, et~al., {Reconstruction of cosmic ray air showers with
  Tunka-Rex data using template fitting of radio pulses}, Phys. Rev. D 97~(12)
  (2018) 122004.
\newblock \href {http://arxiv.org/abs/1803.06862} {\path{arXiv:1803.06862}},
  \href {http://dx.doi.org/10.1103/PhysRevD.97.122004}
  {\path{doi:10.1103/PhysRevD.97.122004}}.

\bibitem{AERA2018}
A.~Aab, et~al., {Observation of inclined {EeV} air showers with the radio
  detector of the Pierre Auger Observatory}, Journal of Cosmology and
  Astroparticle Physics 2018~(10) (2018) 026--026.
\newblock \href {http://dx.doi.org/10.1088/1475-7516/2018/10/026}
  {\path{doi:10.1088/1475-7516/2018/10/026}}.

\bibitem{Schroder:2016hrv}
F.~G. {Schr\"oder}, {Radio detection of Cosmic-Ray Air Showers and High-Energy
  Neutrinos}, Progress in Particle and Nuclear Physics 93 (2017) 1--68.
\newblock \href {http://arxiv.org/abs/1607.08781} {\path{arXiv:1607.08781}}.

\bibitem{Huege:2016veh}
T.~Huege, {Radio detection of cosmic ray air showers in the digital era},
  Physical Reports 620 (2016) 1--52.
\newblock \href {http://arxiv.org/abs/1601.07426} {\path{arXiv:1601.07426}}.

\bibitem{Bechtol:2021tyd}
K.~Bechtol, et~al., {SLAC T-510 experiment for radio emission from particle
  showers: Detailed simulation study and interpretation}, Phys. Rev. D 105~(6)
  (2022) 063025.
\newblock \href {http://arxiv.org/abs/2111.04334} {\path{arXiv:2111.04334}},
  \href {http://dx.doi.org/10.1103/PhysRevD.105.063025}
  {\path{doi:10.1103/PhysRevD.105.063025}}.

\bibitem{T-510:2015pyu}
K.~Belov, et~al., {Accelerator measurements of magnetically-induced radio
  emission from particle cascades with applications to cosmic-ray air showers},
  Phys. Rev. Lett. 116~(14) (2016) 141103.
\newblock \href {http://arxiv.org/abs/1507.07296} {\path{arXiv:1507.07296}},
  \href {http://dx.doi.org/10.1103/PhysRevLett.116.141103}
  {\path{doi:10.1103/PhysRevLett.116.141103}}.

\bibitem{Alvarez_Mu_iz_2018}
J.~Alvarez-Mu{\~{n}}iz, W.~R. Carvalho, K.~Payet, A.~Romero-Wolf,
  H.~Schoorlemmer, E.~Zas, {Comprehensive approach to tau-lepton production by
  high-energy tau neutrinos propagating through the Earth}, Phys. Rev. D
  97~(2).
\newblock \href {http://dx.doi.org/10.1103/physrevd.97.023021}
  {\path{doi:10.1103/physrevd.97.023021}}.

\bibitem{Bertou_2002}
X.~Bertou, P.~Billoir, O.~Deligny, C.~Lachaud, A.~Letessier-Selvon, {Tau
  neutrinos in the Auger Observatory: a new window to {UHECR} sources},
  Astroparticle Physics 17~(2) (2002) 183--193.
\newblock \href {http://dx.doi.org/10.1016/s0927-6505(01)00147-5}
  {\path{doi:10.1016/s0927-6505(01)00147-5}}.

\bibitem{Romero-Wolf:2020pzh}
A.~Romero-Wolf, et~al., {An Andean Deep-Valley Detector for High-Energy Tau
  Neutrinos}, in: {Latin American Strategy Forum for Research Infrastructure},
  2020.
\newblock \href {http://arxiv.org/abs/2002.06475} {\path{arXiv:2002.06475}}.

\bibitem{Ahnen_2018}
M.~Ahnen, et~al., {Limits on the flux of tau neutrinos from 1~{PeV} to 3~{EeV}
  with the {MAGIC} telescopes}, Astroparticle Physics 102 (2018) 77--88.
\newblock \href {http://dx.doi.org/10.1016/j.astropartphys.2018.05.002}
  {\path{doi:10.1016/j.astropartphys.2018.05.002}}.

\bibitem{Brown:2021tf}
A.~Brown, M.~Bagheri, M.~Doro, E.~Gazda, D.~Kieda, C.~Lin, N.~Otte, I.~Taboada,
  A.~Wang, {Trinity: an imaging air Cherenkov telescope to search for
  Ultra-High-Energy neutrinos.}, PoS ICRC2021 (2021) 1179.
\newblock \href {http://dx.doi.org/10.22323/1.395.1179}
  {\path{doi:10.22323/1.395.1179}}.

\bibitem{Venters_2020}
T.~M. Venters, M.~H. Reno, J.~F. Krizmanic, L.~A. Anchordoqui, C.~Gu{\'{e}}pin,
  A.~V. Olinto, {POEMMA}'s target-of-opportunity sensitivity to cosmic neutrino
  transient sources, Phys. Rev. D 102~(12).
\newblock \href {http://dx.doi.org/10.1103/physrevd.102.123013}
  {\path{doi:10.1103/physrevd.102.123013}}.

\bibitem{GRAND:2018iaj}
J.~\'Alvarez-Mu\~niz, et~al., {The Giant Radio Array for Neutrino Detection
  (GRAND): Science and Design}, Sci. China Phys. Mech. Astron. 63~(1) (2020)
  219501.
\newblock \href {http://arxiv.org/abs/1810.09994} {\path{arXiv:1810.09994}},
  \href {http://dx.doi.org/10.1007/s11433-018-9385-7}
  {\path{doi:10.1007/s11433-018-9385-7}}.

\bibitem{FLIESCHER2012S124}
S.~Fliescher, {Radio detection of cosmic ray induced air showers at the Pierre
  Auger Observatory}, Nuclear Instruments and Methods in Physics Research
  Section A: Accelerators, Spectrometers, Detectors and Associated Equipment
  662 (2012) S124--S129, 4th International workshop on Acoustic and Radio EeV
  Neutrino detection Activities.
\newblock \href {http://dx.doi.org/10.1016/j.nima.2010.11.045}
  {\path{doi:10.1016/j.nima.2010.11.045}}.

\bibitem{PierreAuger:2015aqe}
A.~Aab, et~al., {Nanosecond-level time synchronization of autonomous radio
  detector stations for extensive air showers}, Journal of Instrumentation
  11~(01) (2016) P01018--P01018.
\newblock \href {http://dx.doi.org/10.1088/1748-0221/11/01/p01018}
  {\path{doi:10.1088/1748-0221/11/01/p01018}}.

\bibitem{TAROGE}
J.~W. {Nam}, C.~C. {Chen}, C.~H. {Chen}, C.~W. {Chen}, P.~{Chen}, Y.~C. {Chen},
  S.~Y. {Hsu}, J.~J. {Huang}, M.~H.~A. {Huang}, T.~C. {Liu},
  J.~{{\v{R}}{\'\i}pa}, Y.~S. {Shiao}, M.~Z. {Wang}, S.~H. {Wang}, {Design and
  implementation of the TAROGE experiment}, International Journal of Modern
  Physics D 25~(13) (2016) 1645013.
\newblock \href {http://dx.doi.org/10.1142/S0218271816450139}
  {\path{doi:10.1142/S0218271816450139}}.

\bibitem{Prechelt_2022}
R.~Prechelt, et~al., {Analysis of a tau neutrino origin for the near-horizon
  air shower events observed by the fourth flight of the Antarctic Impulsive
  Transient Antenna}, Phys. Rev. D 105~(4).
\newblock \href {http://dx.doi.org/10.1103/physrevd.105.042001}
  {\path{doi:10.1103/physrevd.105.042001}}.

\bibitem{PUEO_white}
Q.~Abarr, et~al., {The Payload for Ultrahigh Energy Observations ({PUEO}): a
  white paper}, Journal of Instrumentation 16~(08) (2021) P08035.
\newblock \href {http://dx.doi.org/10.1088/1748-0221/16/08/p08035}
  {\path{doi:10.1088/1748-0221/16/08/p08035}}.

\bibitem{Abraham:2022jse}
R.~M. Abraham, et~al., {Tau Neutrinos in the Next Decade: from GeV to EeV
  }\href {http://arxiv.org/abs/2203.05591} {\path{arXiv:2203.05591}}, \href
  {http://dx.doi.org/10.48550/ARXIV.2203.05591}
  {\path{doi:10.48550/ARXIV.2203.05591}}.

\bibitem{Vieregg_2016}
A.~Vieregg, K.~Bechtol, A.~Romero-Wolf, {A technique for detection of {PeV}
  neutrinos using a phased radio array}, Journal of Cosmology and Astroparticle
  Physics 2016~(02) (2016) 005--005.
\newblock \href {http://dx.doi.org/10.1088/1475-7516/2016/02/005}
  {\path{doi:10.1088/1475-7516/2016/02/005}}.

\bibitem{V:2019gld}
A.~B. V., A.~Haungs, T.~Huber, T.~Huege, M.~Kleifges, M.~Renschler,
  H.~Schieler, F.~G. Schr\"oder, A.~Weindl, {A Surface Radio Array for the
  Enhancement of IceTop and its Science Prospects}, EPJ Web Conf. 216 (2019)
  04004.
\newblock \href {http://arxiv.org/abs/1907.04171} {\path{arXiv:1907.04171}},
  \href {http://dx.doi.org/10.1051/epjconf/201921604004}
  {\path{doi:10.1051/epjconf/201921604004}}.

\bibitem{hughes2019towards}
K.~Hughes, et~al., {Towards Interferometric Triggering on Air Showers Induced
  by Tau Neutrino Interactions}, PoS (ICRC2019) 917.
\newblock \href {http://dx.doi.org/10.22323/1.358.0917}
  {\path{doi:10.22323/1.358.0917}}.

\bibitem{LWA_OVRO}
M.~W. {Eastwood}, M.~M. {Anderson}, R.~M. {Monroe}, G.~{Hallinan}, B.~R.
  {Barsdell}, S.~A. {Bourke}, M.~A. {Clark}, S.~W. {Ellingson}, J.~{Dowell},
  H.~{Garsden}, {The Radio Sky at Meter Wavelengths: m-mode Analysis Imaging
  with the OVRO-LWA}, Astron. J. 156~(1) (2018) 32.
\newblock \href {http://arxiv.org/abs/1711.00466} {\path{arXiv:1711.00466}},
  \href {http://dx.doi.org/10.3847/1538-3881/aac721}
  {\path{doi:10.3847/1538-3881/aac721}}.

\bibitem{Ellingson20091421}
S.~W. Ellingson, T.~E. Clarke, A.~Cohen, J.~Craig, N.~E. Kassim, Y.~Pihlstrom,
  L.~J. Rickard, G.~B. Taylor, {The Long Wavelength Array}, Proceedings of the
  IEEE 97~(8) (2009) 1421 – 1430.
\newblock \href {http://dx.doi.org/10.1109/JPROC.2009.2015683}
  {\path{doi:10.1109/JPROC.2009.2015683}}.

\bibitem{voors20154nec2}
A.~Voors, \href{http://https://www.qsl.net/4nec2/}{{4nec2, NEC based antenna
  modeler and optimizer}} (2015).
\newline\urlprefix\url{http://https://www.qsl.net/4nec2/}

\bibitem{xfdtd}
{Remcom}, \href{https://www.remcom.com/xfdtd-3d-em-simulation-software}{{XFdtd
  3D Electromagnetic Simulation Software}}.
\newline\urlprefix\url{https://www.remcom.com/xfdtd-3d-em-simulation-software}

\bibitem{Apel:2012uyt}
W.~D. Apel, et~al., {LOPES-3D, an antenna array for full signal detection of
  air-shower radio emission}, Nucl. Instrum. Meth. A 696 (2012) 100--109.
\newblock \href {http://arxiv.org/abs/1303.6808} {\path{arXiv:1303.6808}},
  \href {http://dx.doi.org/10.1016/j.nima.2012.08.082}
  {\path{doi:10.1016/j.nima.2012.08.082}}.

\bibitem{Charrier:2015lsa}
D.~Charrier, {Design of a low noise, wide band, active dipole antenna for a
  cosmic ray radiodetection experiment}\href {http://arxiv.org/abs/1508.02956}
  {\path{arXiv:1508.02956}}, \href {http://dx.doi.org/10.1109/APS.2007.4396539}
  {\path{doi:10.1109/APS.2007.4396539}}.

\bibitem{Ellingson_radioarrays}
S.~Ellingson, {Antennas for the Next Generation of Low-Frequency Radio
  Telescopes}, IEEE Transactions on Antennas and Propagation 53~(8) (2005)
  2480--2489.
\newblock \href {http://dx.doi.org/10.1109/TAP.2005.852281}
  {\path{doi:10.1109/TAP.2005.852281}}.

\bibitem{2001A&A...365..294D}
G.~A. {Dulk}, W.~C. {Erickson}, R.~{Manning}, J.~L. {Bougeret}, {Calibration of
  low-frequency radio telescopes using the galactic background radiation}, AAP
  365 (2001) 294--300.
\newblock \href {http://dx.doi.org/10.1051/0004-6361:20000006}
  {\path{doi:10.1051/0004-6361:20000006}}.

\bibitem{opensky}
{Sch\"{a}fer, Matthias and Strohmeier, Martin and Lenders, Vincent and
  Martinovic, Ivan and Wilhelm, Matthias },
  \href{https://opensky-network.org/}{{Bringing up OpenSky: A large-scale ADS-B
  sensor network for research}} (Apr 2014).
\newline\urlprefix\url{https://opensky-network.org/}

\bibitem{allison2019design}
P.~Allison, et~al., {Design and Performance of an Interferometric Trigger Array
  for Radio Detection of High-Energy Neutrinos}, Nuclear Instruments and
  Methods in Physics Research Section A: Accelerators, Spectrometers, Detectors
  and Associated Equipment 930 (2019) 112--125.
\newblock \href {http://dx.doi.org/https://doi.org/10.1016/j.nima.2019.01.067}
  {\path{doi:https://doi.org/10.1016/j.nima.2019.01.067}}.

\bibitem{Zeolla:2021bl}
A.~Zeolla, S.~A. Wissel, J.~Alvarez-Mu\~{n}iz, W.~Carvalho~Jr., A.~C. Cummings,
  Z.~Curtis-Ginsberg, C.~Deaconu, K.~Hughes, A.~Ludwig, K.~Mulrey, E.~Oberla,
  S.~Prohira, A.~Romero-Wolf, H.~Schoorlemmer, D.~Southall, A.~Vieregg, E.~Zas,
  {Modeling and Validating RF-Only Interferometric Triggering with Cosmic Rays
  for BEACON}, PoS ICRC2021 (2021) 1072.
\newblock \href {http://dx.doi.org/10.22323/1.395.1072}
  {\path{doi:10.22323/1.395.1072}}.

\bibitem{rtk}
Y.~Feng, J.~Wang, {GPS RTK Performance Characteristics and Analysis}, Journal
  of Global Positioning Systems 7.
\newblock \href {http://dx.doi.org/10.5081/jgps.7.1.1}
  {\path{doi:10.5081/jgps.7.1.1}}.

\bibitem{zed-f9p-manual}
{u-blox}, {ZED-F9P, u-blox F9 high precision GNSS module}, {u-blox}.

\bibitem{unavco}
{UNAVCO Community}, {PBO GPS Network - P652 - BarcroftObCS2007} (2007).
\newblock \href {http://dx.doi.org/10.7283/T5QC01GT}
  {\path{doi:10.7283/T5QC01GT}}.

\bibitem{Romero-Wolf:2014pua}
A.~Romero-Wolf, et~al., {An interferometric analysis method for radio impulses
  from ultra-high energy particle showers}, Astropart. Phys. 60 (2015) 72--85.
\newblock \href {http://dx.doi.org/10.1016/j.astropartphys.2014.06.006}
  {\path{doi:10.1016/j.astropartphys.2014.06.006}}.

\bibitem{southallmoriond}
D.~Southall, et~al., {Isolating Cosmic Ray Candidates with the BEACON
  Prototype}, Contribution to the 2022 Very High Energy Phenomena in the
  Universe session of the 56th Rencontres de Moriond.

\bibitem{PhysRevD.93.122005}
A.~Aab, et~al., {Energy estimation of cosmic rays with the Engineering Radio
  Array of the Pierre Auger Observatory}, Phys. Rev. D 93 (2016) 122005.
\newblock \href {http://dx.doi.org/10.1103/PhysRevD.93.122005}
  {\path{doi:10.1103/PhysRevD.93.122005}}.

\bibitem{gorham2018constraints}
P.~Gorham, et~al., {Constraints on the diffuse high-energy neutrino flux from
  the third flight of ANITA}, Phys. Rev. D 98~(2) (2018) 022001.
\newblock \href {http://dx.doi.org/10.1103/PhysRevD.98.022001}
  {\path{doi:10.1103/PhysRevD.98.022001}}.

\bibitem{zilles2020radio}
A.~Zilles, et~al., {Radio Morphing: towards a fast computation of the radio
  signal from air showers}, Astroparticle Physics 114 (2020) 10--21.
\newblock \href {http://dx.doi.org/10.1016/j.astropartphys.2019.06.001}
  {\path{doi:10.1016/j.astropartphys.2019.06.001}}.

\bibitem{CFR-ADS-B-91.225}
{Office of the Federal Register, National Archives and Records Administration},
  \href{https://www.ecfr.gov/current/title-14/chapter-I/subchapter-F/part-91/subpart-C/section-91.225}{{14
  CFR \S 91.225 - Automatic Dependent Surveillance-Broadcast (ADS-B) Out
  equipment and use}} (December 2020).
\newline\urlprefix\url{https://www.ecfr.gov/current/title-14/chapter-I/subchapter-F/part-91/subpart-C/section-91.225}

\bibitem{CFR-ADS-B-91.227}
{Office of the Federal Register, National Archives and Records Administration},
  \href{https://www.ecfr.gov/current/title-14/chapter-I/subchapter-F/part-91/subpart-C/section-91.227}{{14
  CFR \S 91.227 Automatic Dependent Surveillance-Broadcast (ADS-B) Out
  equipment performance requirements}} (December 2020).
\newline\urlprefix\url{https://www.ecfr.gov/current/title-14/chapter-I/subchapter-F/part-91/subpart-C/section-91.227}

\bibitem{charrier2019autonomous}
D.~Charrier, et~al., {Autonomous radio detection of air showers with the
  TREND50 antenna array}, Astroparticle Physics 110 (2019) 15--29.

\bibitem{Nam:2021gcb}
J.~Nam, et~al., {Development of drone-borne aerial calibration pulser system
  for radio observatories of ultra-high energy air showers}, PoS ICRC2021
  (2021) 283.
\newblock \href {http://dx.doi.org/10.22323/1.395.0283}
  {\path{doi:10.22323/1.395.0283}}.

\bibitem{Mulrey:2021fqw}
K.~Mulrey, {Cross-calibrating the energy scales of cosmic-ray experiments using
  a portable radio array}, PoS ICRC2021 (2021) 414.
\newblock \href {http://dx.doi.org/10.22323/1.395.0414}
  {\path{doi:10.22323/1.395.0414}}.

\end{thebibliography}

\end{document}